\documentclass[hdvipdfm]{ptephy_v1}

\preprintnumber{XXXX-XXXX} 

\usepackage{braket}
\usepackage{caption}
\usepackage{here}

\usepackage{here}
\def\slash#1{\not\!#1}
\def\slashb#1{\not\!\!#1}

\begin{document}
\title{$K^{+}$-nucleus elastic scattering revisited from perspective of partial restoration of chiral symmetry}


%
  
 \author{
  \name{\fname{Kenji} \surname{Aoki}}{\ast} and
  \name{\fname{Daisuke} \surname{Jido}}{}}
\address{
  Department of Physics, Tokyo Metropolitan University, Hachioji, Tokyo 192-0397, Japan \email{aoki-kenji1@ed.tmu.ac.jp} 
}

%
%
%


\begin{abstract}
The $K^{+}$ meson properties in the nuclear medium are investigated
by considering the wavefunction renormalization
as a first step to reveal 
the in-medium properties of the $K^{+}$ meson
in the context of partial restoration of chiral symmetry.
The $K^{+}N$ elastic scattering amplitude is constructed using chiral perturbation theory
up to the next-to-leading order.
Using the constructed amplitude, we calculate the wavefunction renormalization
in the Thomas-Fermi approximation.
We obtained a good description of the $K^{+}N$ elastic scattering amplitude.
The obtained wavefunction renormalization factor suggests the 2 to 6\% enhancement
of the $K^{+}N$ interaction at the saturation density.
We conclude that the wavefunction renormalization could be one of the
important medium effects for the $K^{+}$ meson.
\end{abstract}

\subjectindex{D33, D32, B60, B69}

\maketitle
\section{Introduction}
\label{sec:1}
The strong interaction properties of hadron in the nuclear medium has attracted considerable interests,
in particular, for the purpose of proving the spontaneous breaking of chiral symmetry from the experimental facts.
The spontaneous breaking of chiral symmetry is considered to be one of the phase transition phenomena in
the Quantum Chromodynamics (QCD) vacuum.
Thus, by changing the temperature and/or density of the system, 
one expects that the broken symmetry might be restored.
With this nature, we try to prove the mechanism of the spontaneous breaking of chiral symmetry from the experimental facts.
In the phase transition of the QCD vacuum, the quark condensate 
$\langle \bar{q}q \rangle$ is one of the order parameters that characterize chiral symmetry breaking.
At high temperature and/or high density, the magnitude of the quark condensate is considered to be decreasing.
Especially, in the finite density and zero temperature system, such as atomic nuclei, 
chiral symmetry is considered to be
partially restored and the magnitude of the quark condensate is sufficiently reduced.
So, we study the partial restoration of chiral symmetry, which may be more easily accessible situation than 
the complete restoration of chiral symmetry at extreme conditions,
in order to prove the mechanism of the spontaneous breaking of chiral symmetry. 
For this purpose, we study the behavior of the quark condensate in the nuclear medium.
However, since the quark condensate is not a direct observable quantity, 
one is forced to extract  the information of the quark condensate indirectly
by analyzing the experimental observables, 
such as hadron-nucleus scattering. 
Suitable systems for such studies may be Nambu-Goldstone (NG) bosons in the nuclear medium, because the NG bosons
should be sensitive to the spontaneous breaking of chiral symmetry and the NG bosons change their properties in nuclear medium.
Comparing in-medium and in-vacuum NG boson properties quantitatively,
one may prove the reduction of the magnitude of the quark condensate quantitatively from the observations.

The study of the partial restoration of chiral symmetry have been carried out especially for pion.
From the observations of the deeply bound pionic atom \cite{ks} and low-energy pion-nucleus elastic scattering \cite{f2004}
taking into accounts the theoretical considerations \cite{k2003, j2008}, 
chiral symmetry is considered to be restored about 30\% at the saturation density.
So, systematic studies for other systems are necessary, and we need to confirm consistency of partial restoration of 
chiral symmetry in other systems.
Yet, we do not know whether the partial restoration of chiral symmetry systematically occurs
in other systems than the pion-nucleus system.
For this purpose, we would like to consider the NG bosons with strangeness.

The promising proves to investigate the nature of the strangeness in nuclear medium are $K^{\pm}$.
It is known that the strong interaction properties of $K^{+}$ and $K^{-}$ with nucleon, or with nuclei,
are considerably different due to their strangeness content \cite{review}.
It is assigned that $S=+1$ for $K^+$ and $S=-1$ for $K^{-}$.
We know that $\bar{K}N$ interaction is so strong attractive 
that $\Lambda(1405)$ has been considered to appear as a quasi bound state of $\bar{K}N$ \cite{lambda}.
When we consider the $\bar{K}$-nucleus system, 
strong absorption into the nucleus masks medium effects on the $\bar{K}$ meson.
On the other hand, 
the $K^{+}N$ interaction in vacuum is repulsive and relatively weak in the low-energy region 
$p_{{\rm lab}} \leq$ 800 MeV/c where inelasticity is small.
The $K^{+}N$ cross section is small in comparison with $K^{-}N$ and
the mean free path in nuclear medium is long and comparable 
to the typical nuclear size 5-7 fm.
Further, there is no hyperon resonances strongly coupled to $K^{+}N$.
With this nature, the $K^{+}$ meson can penetrate deeply in the nucleus
without suffering from the strong absorption.
Therefore, it is suggested that the $K^{+}$ meson is capable of using as good probe to investigate
the nature of the strangeness in nuclear medium.

Based on the fact that the mean free path of the $K^{+}$ meson in nuclear medium is comparable
to the typical nuclear size, one would consider that the $K^{+}$ meson scattering with nucleons in the nucleus
could be described by the single step scattering, that is the linear density approximation would be valid.
It is remarkable that the ratio $R$ of the total elastic cross section for $K^{+}$-carbon on
the total elastic cross section for $K^{+}$-deuteron per nucleon is given by
\begin{eqnarray}
R = \frac{\sigma(K^{+} \thinspace ^{12}{\rm C})}{6 \sigma(K^{+} \thinspace d)} > 1
\label{eq:ratio}
\end{eqnarray}
in the region of kaon lab momenta $p_{{\rm lab}}$ = 450-900 MeV/c \cite{ratio1, ratio2, krauss}.
By the expectation of the in-vacuum $K^{+}N$ interaction, the $K^{+}$-carbon scattering would satisfy
the linear density approximation, namely $\sigma(K^{+} \thinspace ^{12} {\rm C}) \simeq 12 \sigma(K^{+} N)$.
However, Eq. (\ref{eq:ratio}) suggests that the single-step scattering does not explain 
the elastic total cross section for $K^{+}$-carbon scattering.
Even, the ratio should be expected as rather $R<1$ if one considers
the nuclear shadowing effect.
This observation was also found as breakdown of 
the low-density $T\rho$ approximation in the $K^+$ optical potential 
in nuclei extracted from data of $K^+$-nucleus elastic scatterings \cite{f2007}. 
The study of Ref. \cite{f2007} tells us that the optical potential 
which reproduces the experimental data for the $K^{+}$-nucleus elastic scattering is 
14-34\% enhanced than what one expects from a simple $T\rho$ approximation. 

To understand such unanticipated enhancement of the $K^+N$ interaction in nuclear medium,
several ideas have been proposed.
As unconventional in-medium effects, 
the ``swelling" of nucleons in nuclei was firstly suggested in Ref. \cite{swell} and
later in Ref. \cite{peterson}.
The model which could provide the physical interpretation of the 
``swelling" was proposed in Ref. \cite{vec} considering 
the reduction of the mass of the vector meson, which intermediates 
the effective $K^{+}N$ interaction, in the nuclear medium.
Other possibilities were discussed, for instance, 
by considering the medium corrections on the meson exchange current \cite{jiang1992}, 
by including pion cloud contribution \cite{garcia1995}
and by putting medium effects on exchanged mesons 
between the $K^+$ and the target nucleons \cite{caillon1996}.
The recent studies using the optical potential based on the in-medium kinematics 
were done in Ref. \cite{f2016, meson2016}.

In this paper, we would like to describe the enhancement of the $K^+ N$ interaction 
in the nuclear medium in terms of the wavefunction renormalization of the in-medium kaon \cite{j2016}. 
The wavefunction renormalization should be taken into account, 
if the self-energy has energy-dependence. 
The self-energy for the NG boson is expected to have substantial energy dependence, 
because, according to the chiral effective theory, 
the amplitude is written in terms of the derivative expansion of the NG boson field. 
Thus, we expect that the wavefunction renormalization plays a crucial role 
as one of the leading-order corrections to the $K^{+}N$ interaction. 
In addition, the wavefunction renormalization is closely connected 
to partial restoration of chiral symmetry \cite{k2003, j2001, j2008}, 
which provides one of the fundamental interpretations of the in-medium properties of hadron. 
In the chiral effective theory, the NG boson field is introduced as a parametrization of the coset space of the broken symmetry, 
and thus should be normalized by a dimensional quantity because the physical boson field has mass dimension 
while the field parametrizing the coset space is dimensionless. 
Once partial restoration of chiral symmetry takes places in nuclear matter by reduction of the chiral order parameters, 
it changes the energy scale of the vacuum and renormalizes the NG boson field. 
For these reasons, we would like to examine the wavefunction renormalization of the in-medium $K^{+}$ 
using the in-vacuum $K^{+}N$ elastic scattering amplitude calculated 
from chiral perturbation theory and its unitarization. 
This study can be a first step to describe the in-medium $K^{+}$ property 
by more systematic approaches, such as in-medium chiral perturbation theory. 

This paper is organized as follows.
In Sec. \ref{sec:2}, we summarize the model of the in-medium kaon 
self-energy and investigate the properties of the wavefunction renormalization.
In Sec. \ref{sec:3}, we describe the formulation of chiral perturbation theory and 
calculate the in-vacuum $K^{+}N$ amplitude.
In Sec. \ref{sec:4}, we discuss the numerical results.
The total cross section and differential cross section of the $K^{+}N$ elastic scattering are presented.
We discuss the wavefunction renormalization.
In Sec. \ref{sec:5}, we conclude the results of this paper.

\section{Wavefunction renormalization}
\label{sec:2}
As stated in introduction, the aim of this paper is to describe the role of the wavefunction renormalization 
in the in-medium $K^{+}$ self-energy, that is the optical potential for $K^{+}$ in the nuclear medium. 
For this purpose, we take a simple description of the $K^{+}$ self-energy based on the $K^{+}N$ elastic scattering amplitude:
\begin{equation}
  \Pi(\omega, \vec p) =4 \int^{k_{F}} \frac{d^{3} q}{(2\pi)^{3} 2M_{N}} T_{K^{+}N}(p,q) \label{eq:selfene}
\end{equation}
where $T_{K^{+}N}(p,q)$ is the forward elastic $K^{+}N$ scattering amplitude 
with kaon momentum $p$ and nucleon momentum $q$ and 
is integrated with respect to the nucleon spacial momentum 
up to the Fermi momentum $k_{F}=(3\pi^{2} \rho/2)^{1/3}$. 
Here we consider the isospin symmetric nuclear matter. 
The factor 4 is the multiplicity of the spin and isospin.
The bound nucleons in the nuclear medium are 
described by Thomas-Fermi approximation in which the nucleons are treated as Fermi gas 
in a potential $v_{N} = - k_{F}^{2}/2M_{N}$ and the energy of the nucleon 
with momentum $q$ is given by $E_{N}=M_{N}+ q^{2}/2M_{N} + v_{N}$.   
Here the nucleon spinor is normalized as $\bar u(q,s) u(q,s^{\prime}) = 2M_{N} \delta_{ss^{\prime}}$. 
The scattering amplitude $T_{K^{+}N}(p,q)$ is evaluated in the nuclear matter rest frame
and is allowed to extend to the energy region off the mass shell of $K^{+}$ 
according to chiral perturbation theory. 
The off-shell extension is necessary to calculate the wavefunction renormalization, 
which is obtained by partial derivative of the self-energy with respect to the $K^{+}$ energy 
by fixing the momentum. 
The description of the $K^{+}N$ scattering amplitude will be given 
in the next section. 
If one neglects Fermi motion of the nucleons in the nuclear matter, 
one finds that Eq.~(\ref{eq:selfene}) is reduced to the $T\rho$ approximation 
\begin{equation}
\Pi(\omega, \vec p) =  \rho T_{K^{+}N}(\omega, \vec p).  \label{eq:Trho}
\end{equation}

Let us consider the case that the $K^{+}$ self-energy in the nuclear medium should have energy dependence. 
In such a case, the effect of the energy dependence can be implemented to an equivalent energy independent self-energy at the $K^{+}$ pole with the wavefunction renormalization factor as follows. Following the discussion for the $\pi$ meson in Ref. \cite{k2003}, we introduce the Klein-Gordon equation for the in-medium $K^{+}$ with energy $\omega^{*}$, momentum $\vec p$ and the in-vacuum mass $M_{K}$ as
\begin{equation}
   \omega^{*2} - \omega^{2} - \Pi(\omega^{*},\vec p) =0  \label{eq:KGeq}
\end{equation}
with $\omega = \sqrt{\vec p^{\,2} + M^{2}_{K}}$ and the self-energy $\Pi(\omega^{*},\vec p)$. Here let us assume that $\Pi(\omega^{*}) \ll \omega^{2}$, which means that the one-body potential for $K^{+}$ produced by the nuclear medium is sufficiently small in comparison with the free kaon energy, so that the effects of the self-energy to the free $K^{+}$ meson is considered to be moderately small, $\omega^{*} \simeq \omega$. With this assumption, expanding the self-energy around $\omega^{*} = \omega$ and evaluating it at the on-shell condition Eq. (\ref{eq:KGeq}) with $\omega^{*}=\omega$, we obtain
\begin{eqnarray}
  \Pi(\omega^{*}) &=& \Pi(\omega) + (\omega^{*2} - \omega^{2}) \left.\frac{\partial \Pi}{\partial \omega^{*2}} \right|_{\omega^{*}=\omega} + \cdots 
  \nonumber \\
  &\simeq& \left( 1+  \left.\frac{\partial \Pi}{\partial \omega^{*2}} \right|_{\omega^{*}=\omega} \right) \Pi(\omega)
  \equiv Z \Pi(\omega),  \label{eq:renormalization}\\
  Z &\equiv& 1+  \left.\frac{\partial \Pi}{\partial \omega^{*2}} \right|_{\omega^{*}=\omega}
   \label{eq:z_factor}
\end{eqnarray}
where we neglect the higher orders of $(\omega^{*2}-\omega^{2})$ and in the second line we assume that the difference between $\Pi(\omega^{*})$ and $\Pi(\omega)$ is in higher orders in density expansion. Here we have introduced the wavefunction renormalization factor as $Z$. Therefore, when the self-energy has sufficiently strong energy-dependence, which may be the case for the NG bosons, the wavefunction renormalization plays an important role as one of the substantial in-medium effects. 

In the present model, the self-energy is given by the $K^{+}N$ scattering amplitude as shown in Eq.~(\ref{eq:selfene}). 
Thus, the wavefunction renormalization factor $Z$ is obtained from the $K^{+}$ energy dependence of the scattering amplitude. 
In the leading order of the density expansion, the in-medium self-energy 
is given by the $T\rho$ approximation (\ref{eq:Trho}). 
Thus, the wavefunction renormalization is one of the next-to-leading corrections 
for the in-medium self-energy. (The leading correction is given by the in vacuum $K^{+}N$ scattering amplitude.) 
Here we would like to emphasize that the wavefunction renormalization 
is the leading order correction to the $K^{+}N$ interaction. In the following sections, 
we describe the $K^{+}N$ scattering amplitude based on 
chiral perturbation theory and see whether the wavefunction renormalization explains 
the enhancement of the $K^{+}N$ interaction in the nuclear medium.

\section{Formulation for $K^{+}N$ amplitudes}
\label{sec:3}
In this section we describe the $KN$ elastic scattering amplitude. 
We calculate the $KN$ scattering amplitude at the tree level based 
on chiral perturbation theory up to the next-to-leading order. 
We determine the low energy constants appearing in the next-to-leading order 
so as to reproduce the observed $K^{+}p$ differential cross section at certain momentum
and $I=0$ total cross section.
To compare the calculated amplitude with the observation, 
we consider the Coulomb correction, 
which becomes important especially for the forward scattering, to the $K^{+}p$ amplitude. 

\subsection{Chiral perturbation theory}
We use chiral perturbation theory as a low energy effective theory of QCD
to describe the $K^{+}N$ interaction.
Chiral perturbation theory which has $SU(3)_{L} \times SU(3)_{R}$ symmetry describes interaction between 
the pseudoscalar NG bosons and ${\frac{1}{2}}^{+}$ baryon octet.
The leading order chiral Lagrangian is given for the baryon field $B$ as
\begin{eqnarray}
{\cal L}_{MB}^{(1)}=
{\rm Tr} \left[\bar{B}(i \slashb{D} - M_{0}) B\right]
- \frac{D}{2} {\rm Tr} \left(\bar{B} \gamma_{\mu} \gamma_{5} \{u^{\mu},B\} \right)
-\frac{F}{2} {\rm Tr} \left (\bar{B} \gamma_{\mu} \gamma_{5} [u^{\mu},B] \right)
\label{eq:MB_1}, 
\end{eqnarray}
where $M_{0}$ is the baryon mass at the chiral limit, and
\begin{eqnarray}
&&D_{\mu} B= \partial_{\mu} B + [ \Gamma_{\mu}, B ],  \\
&&\Gamma_{\mu} = \frac{1}{2} ( \xi^{\dagger} \partial_{\mu} \xi + \xi \partial_{\mu} \xi^{\dagger} ),  \\
&&U=\xi^{2}=\exp \left( i \frac{\sqrt{2}}{f} \Phi \right), \\
&&u_{\mu} = i \left(\xi^{\dagger} \partial_{\mu} \xi - \xi \partial_{\mu} \xi^{\dagger} \right).
\end{eqnarray}
The constants $D$ and $F$ give the axial vector couplings of the baryons at the tree level.
The meson and baryon fields form the SU(3) matrix given by
\begin{eqnarray}
\Phi&=&\left(
\begin{array}{ccc}
\frac{\pi^{0}}{\sqrt{2}}+\frac{\eta}{\sqrt{6}} & \pi^{+} &K^{+} \\
\pi^{-} & -\frac{\pi^{0}}{\sqrt{2}}+\frac{\eta}{\sqrt{6}} & K^{0} \\
K^{-}& \bar{K}^{0} &  -\frac{2\eta}{\sqrt{6}}
\end{array}
\right), \\
B&=&\left(
\begin{array}{ccc}
\frac{\Sigma^{0}}{\sqrt{2}}+\frac{\Lambda}{\sqrt{6}} &\Sigma^{+}& p \\
\Sigma^{-} & -\frac{\Sigma^{0}}{\sqrt{2}}+\frac{\Lambda}{\sqrt{6}} & n \\
\Xi^{-}& \Xi^{0} &  -\frac{2\Lambda}{\sqrt{6}}
\end{array}
\right).
\end{eqnarray}
The next-to-leading order chiral Lagrangian reads
\begin{align}
{\cal L}_{MB}^{(2)}
=&b_{D} {\rm Tr} \left(\bar{B} \{ \chi_{+}, B \} \right)
+b_{F} {\rm Tr} \left(\bar{B} [\chi_{+}, B] \right)
+b_{0} {\rm Tr} \left(\bar{B}B) {\rm Tr} (\chi_{+} \right)  \nonumber \\
&+d_{1} {\rm Tr} \left(\bar{B} \{ u_{\mu}, [u^{\mu}, B] \} \right)  
+d_{2} {\rm Tr} \left(\bar{B} [u_{\mu}, [u^{\mu}, B]] \right)
+d_{3} {\rm Tr} \left(\bar{B} u_{\mu} ) {\rm Tr} (u^{\mu} B \right) \nonumber \\
&+d_{4} {\rm Tr} \left(\bar{B} B) {\rm Tr} (u^{\mu} u_{\mu} \right),
\label{eq:MB_2}
\end{align}
with
\begin{eqnarray}
&&\chi_{+} = 2 B_{0} \left (\xi {\cal M} \xi + \xi^{\dagger} {\cal M} \xi^{\dagger} \right),  \\
&&{\cal M}={\rm diag} \left(\hat{m}, \hat{m}, m_{s} \right),
\end{eqnarray}
where $b_{D}, b_{F}, b_{0}, d_{1}, d_{2}, d_{3}$ and $d_{4}$ are the low-energy constants.
We assume the isospin symmetry and $\hat{m} \equiv (m_u + m_d)/2$ is the 
isospin averaged value of the current quark masses. 
The parameter $B_{0}$ is a positive constant connected to the meson mass.
In this work, the parameter $B_0$ is fixed by 
$B_{0}= M_{K} ^{2}/(\hat{m}+m_{s})$ where $M_K$ is the $K$ meson mass.
The low-energy constants in the next-to-leading order chiral Lagrangian are determined by 
the $K^{+}N$ elastic cross section and the baryon masses.
The baryon masses at the next-to-leading order are given by
\begin{eqnarray}
&&M_{\Sigma} = M_{0} -4B_{0} \left[2b_{D}\hat{m} + b_0 (2\hat{m} + m_s) \right], \label{eq:sigma} \\
&&M_{\Xi} = M_{0} -4B_{0} \left[b_{D}(\hat{m}+m_s) + b_F(m_s - \hat{m}) + b_0 (2\hat{m}+m_s)\right], \label{eq:xi}\\
&&M_{N} = M_{0} -4B_{0} \left[b_D (\hat{m}+m_s) + b_F (\hat{m}-m_s) + b_0 (2\hat{m}+m_s)\right], \label{eq:nuc}\\
&&M_{\Lambda} = M_0 -4B_0 \left[b_D \frac{2(\hat{m}+2m_s)}{3} + b_0 (2\hat{m}+m_s) \right] \label{eq:lambda}.
\end{eqnarray}

\label{sec:chpt}
\begin{figure}[t]
 \begin{center}
  \includegraphics[width=100mm]{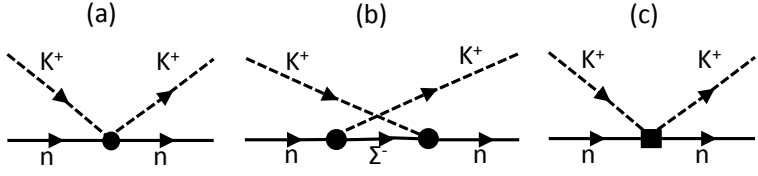}
 \caption{Feynman diagrams for $K^{+}n$ elastic scattering.
 (a) Weinberg-Tomozawa interaction. 
 (b) Crossed Born interaction with the intermediate $\Sigma^{-}$. 
 (c) NLO interaction.}
 \label{fig:n}
 \end{center}
\end{figure}
\begin{figure}[t]
 \begin{center}
  \includegraphics[width=100mm]{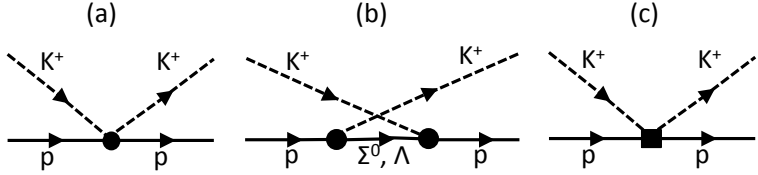}
\caption{Feynman diagrams for $K^{+}p$ elastic scattering.
 (a) Weinberg-Tomozawa interaction. 
 (b) Crossed Born interaction with
 the intermediate $\Sigma^{0}$ or $\Lambda$.
 (c) NLO interaction.}
 \label{fig:p}
 \end{center}
\end{figure}

For the $K^{+}n$ elastic scattering, 
the Feynman diagrams to be calculated are shown Fig. \ref{fig:n}.
The direct Born term does not contributes, 
since there is no baryon resonance with strangeness $S=+1$.
The pentaquark $\Theta^{+}$ can be a candidate for the direct Born term.
But here we do not take into account, because it is known that the
possible $\Theta^{+}$ has very weak coupling to $KN$.
In the crossed Born term, 
the intermediate state has $S=-1$ and $Q=-1$, and the candidate baryon is the $\Sigma^{-}$ baryon.
It turns out to give small contribution.
The $K^{+}n$ invariant amplitudes at the leading order chiral perturbation theory are calculated as
\begin{eqnarray}
&&T_{{\rm WT}}^{K^{+}n} = 
\frac{1}{4f^{2}_{K}}\bar{u}(\vec{p}_4,s_4)( \slash{p_1}+\slash{p_3} )u(\vec{p}_2,s_2), \label{eq:kn1}\\
&&T_{{\rm Born}}^{K^{+}n} 
=-\frac{(D-F)^{2}}{2f^{2}_{K}}\bar{u}(\vec{p}_4,s_4)\slash{p_1}\gamma_5
\frac{M_{\Sigma} + (\slash{p_2} - \slash{p_{3} }) }{M_{\Sigma}^{2} - (p_{2} - p_{3} )^{2} - i\epsilon}
\slash{p_3}\gamma_5u(\vec{p}_2,s_2),
 \label{eq:kn2}
\end{eqnarray}
where the constants $M_{\Sigma}$ and $f_{K}$ 
are the $\Sigma$ baryon mass and kaon decay constant, respectively, with
the initial kaon momentum $p_{1}$, the initial nucleon momentum $p_2$,
the final kaon momentum $p_3$ and the final nucleon momentum $p_4$.
$u(\vec{p}_i, s_i)$ is positive energy Dirac spinor
for the 3-momentum $\vec{p}_{i}$ and spin $s_i$.
Equations (\ref{eq:kn1}) and (\ref{eq:kn2}) are obtained form Weinberg-Tomozawa interaction 
and crossed Born interaction, respectively.
The $K^{+}n$ invariant amplitude at the next-to-leading order chiral perturbation theory is calculated as
\begin{eqnarray}
T_{(2)}^{K^{+}n} = \left[\frac{2B_0}{f^{2}_{K}}(\hat{m}+m_s)(2b_0+b_D-b_F)  
+\frac{2}{f^{2}_{K}}(d_1-d_2-2d_4)(p_1p_3) \right]\bar{u}(\vec{p}_4,s_4)u(\vec{p}_2,s_2).
 \label{eq:kn3}
\end{eqnarray}

For the $K^{+}p$ elastic scattering, 
we calculate the Feynman diagrams shown in Fig. \ref{fig:p}.
Since there is no baryon with $S=+1$ and $Q=2$, we have no direct Born terms.
In the crossed Born term, the $\Sigma^{0}$ baryon and $\Lambda$ baryon
are possible intermediate one-particle states with $S=-1$ and $Q=0$.
The $K^{+}p$ invariant amplitude at the leading order chiral perturbation theory is obtained as
\begin{eqnarray}
T^{K^{+}p}_{{\rm WT}} &=&\frac{1}{2f^{2}_{K}}
\bar{u}(\vec{p}_4,s_4)( \slash{p_1}+\slash{p_3} )u(\vec{p}_2, s_2),   \label{eq:kp1} \\
T^{K^{+}p}_{{\rm Born}} &=& -\left(\frac{D-F}{2f_{K}}\right)^{2}
\bar{u}(\vec{p}_4,s_4)\slash{p_1}\gamma_5
\frac{M_{\Sigma} + (\slash{p_2} - \slash{p_{3} }) }{M_{\Sigma}^{2} - (p_{2} - p_{3} )^{2} - i\epsilon}
\slash{p_3}\gamma_5u(\vec{p}_2,s_2)
\nonumber\\
&&-\left(\frac{3F+D}{2\sqrt{3}f_{K}} \right)^{2}\bar{u}(\vec{p}_4 ,s_4)\slash{p_1}\gamma_5
\frac{M_{\Lambda} + (\slash{p_2} - \slash{p_{3} }) }{M_{\Lambda}^{2} - (p_{2} - p_{3} )^{2} - i\epsilon}
\slash{p_3}\gamma_5u(\vec{p}_2,s_2) \label{eq:kp2}
\end{eqnarray}
with the $\Lambda$ mass $M_{\Lambda}$.
The $K^{+}p$ invariant amplitude at the next-to-leading order chiral perturbation theory 
is calculated as
\begin{eqnarray}
T_{(2)}^{K^{+}p}=
\left[\frac{4B_0}{f^{2}_{K}}(\hat{m}+m_s)(b_0+b_D) 
-\frac{2}{f^{2}_{K}}(2d_2+d_3+2d_4)(p_1p_3) \right]
\bar{u}(\vec{p}_4,s_4)u(\vec{p}_2, s_2).
 \label{eq:kp3}
\end{eqnarray}

The $K^{+}N$ scattering amplitudes in the particle basis, 
$T^{K^{+}p}$ and $T^{K^{+}n}$, 
are decomposed into the amplitude in the isospin basis $T^{I}$ $(I=0, 1)$ as
\begin{eqnarray}
   T^{K^{+}p} &=& T^{I=1}, \\
   T^{K^{+}n} &=& \frac{1}{2} (T^{I=1}+T^{I=0}).
\end{eqnarray}
We work out in the center of mass (c.m.) frame, 
where we perform the partial wave decomposition. 
In the c.m. frame, the invariant amplitude $T^{I}$ is written by the non spin-flip and spin-flip amplitudes, $f^{I}$ and $g^{I}$, as
\begin{equation}
   T^{I}(s, \theta_{\rm c.m.}) = f^{I}(s, \theta_{\rm c.m.}) - i \vec \sigma \cdot \hat n\, g^{I}(s, \theta_{\rm c.m.})
\end{equation}
as functions of the Mandelstam valuable $s$ and the scattering angle $\theta_{\rm c.m.}$ in the c.m.\ frame,
where $\vec\sigma$ is the spin Pauli matrix and the normal vector $\hat n$ of the scattering plane given by
\begin{equation}
  \hat n = \frac{\vec{p}_{3} \times \vec{p}_{1}}{|\vec{p}_{3} \times \vec{p}_{1}|}.
\end{equation}

The differential cross section for the $K^+N$ elastic scattering in the c.m. frame is given by
\begin{eqnarray}
\frac{d \sigma({K^+N})}{d \Omega} =\frac{1}{64 \pi^{2} s} 
\left( | f^{K^+N} |^{2} + | g^{K^+N} |^{2}  \right).
\label{eq:diff}
\end{eqnarray}
The total cross section in the c.m. frame for isospin $I$ is obtained by integrating 
the differential cross section
in terms of the solid angle $d \Omega = \sin\theta_{{\rm c.m.}} d\theta_{{\rm c.m.}} d\phi$
\begin{eqnarray}
\sigma^{I} =  \frac{1}{64 \pi^{2} s} \int d\Omega 
\left(| f^{I} |^{2} + | g^{I} |^{2} \right).
\end{eqnarray}

The amplitudes, $f^{I}$ and $g^{I}$, are expanded into the partial waves with Legendre polynomials $P_{l}(x)$ as
\begin{eqnarray}
  f^{I}(s,\theta_{\rm c.m.}) &=& \sum_{l=0}^{\infty} f_{l}^{I}(s) P_{l}(\cos\theta_{\rm c.m}) 
  \nonumber \\ 
  &=& \sum_{l=0}^{\infty}\left[(l+1) T_{l+}^{I}(s) + l T_{l-}^{I}(s)\right] P_{l}(\cos\theta_{\rm c.m.}), \\
  g^{I}(s,\theta_{\rm c.m.}) &=& \sum_{l=1}^{\infty} g_{l}^{I}(s) \sin\theta_{\rm c.m.} \frac{dP_{l}(\cos\theta_{\rm c.m})}{d\cos\theta_{\rm c.m.}} \nonumber \\  
  &=& \sum_{l=1}^{\infty}\left[T_{l+}^{I}(s) - T_{l-}^{I}(s)\right] \sin\theta_{\rm c.m.} \frac{dP_{l}(\cos\theta_{\rm c.m})}{d\cos\theta_{\rm c.m.}}, 
\end{eqnarray}
where we have introduced the partial wave amplitudes, $T^{I}_{l\pm}(s)$, which have definite total angular momentum. For lower partial waves, the explicit relation of these two amplitudes are found as
\begin{eqnarray}
\label{eq:p1}
&&T_{0+}^{I} = f^{I}_{l=0}, \\
&&T_{1+}^{I} = \frac{1}{3} \left( f^{I}_{l=1} + g^{I}_{l=1} \right ), \\
&&T_{1-}^{I} = \frac{1}{3} \left( f^{I}_{l=1} - 2 g^{I}_{l=1} \right ), \\
&&T_{2+}^{I} = \frac{1}{5} \left( f^{I}_{l=2} + 2g^{I}_{l=2} \right), \\
\label{eq:p2}
&&T_{2-}^{I} = \frac{1}{5} \left(f^{I}_{l=2} - 3 g^{I}_{l=2} \right)
\end{eqnarray} and so on.

As we have seen above, in our formulation, we carry out the partial wave decomposition in the c.m.\ frame.  
The self-energy given in Eq.~(\ref{eq:selfene}), however, 
is calculated in the rest frame of the nuclear medium, in which each nucleon has its own Fermi motion. 
Thus, we make Lorentz transformation from the $K^{+}N$ c.m. frame to the nuclear matter rest frame.

\subsection{Coulomb corrections}
\label{sec:coulomb}
For the $K^{+}p$ scattering, 
we take into account the Coulomb correction to compare the amplitude calculated theoretically 
to the observed differential cross section. 
By following Ref.~\cite{hashimoto}, 
we add the Coulomb amplitude $f_{C}$ to the strong interaction part of 
the $K^{+}p$ scattering multiplied by the Coulomb phase shift factor $e^{2i\Phi_{\ell}}$:
\begin{eqnarray}
&&f^{K^+p}
= \sum_{l=0}^{\infty} \left[ (l+1) T_{l+}^{I=1} + l T_{l-}^{I=1} \right] 
e^{2 i \Phi_{l}} P_{l}(\cos \theta_{{\rm c.m.}}) -8 \pi \sqrt{s} f_{C}, \\
&&g^{K^+p}
= \sum_{l=1}^{\infty} \left[ T_{l+}^{I=1} - T_{l-}^{I=1} \right] e^{2 i \Phi_{l}}   
\sin\theta_{{\rm c.m.}}  \frac{d P_{l}(\cos\theta_{{\rm c.m.}})}{d \cos\theta_{{\rm c.m.}}}.
\end{eqnarray}
The Coulomb amplitude and Coulomb phase shift are described as
\begin{eqnarray}
&&f_{C}=- \frac{\alpha}{2 k v \sin^{2} (\theta/2) } \exp \left[ -i \frac{\alpha}{v} \ln \left(\sin^{2} \frac{\theta}{2} \right) \right], \\
&&\Phi_{l}=\sum_{n=1}^{l} \tan^{-1} \frac{\alpha}{n v}
\end{eqnarray}
for $l > 0$ ($\Phi_{0}$=0) with the relative velocity between the kaon and proton given by
\begin{eqnarray}
v = \frac{k (E_{K} + E_{p} ) } {E_{K} E_{p}},
\end{eqnarray}
where $E_{K}$, $E_{p}$ and $k$ are the kaon energy, proton energy and kaon momentum in the c.m. system,
respectively and $\alpha$ is the fine structure constant.

\section{Numerical result}
\label{sec:4}
\subsection{Tree level $K^{+}N$ amplitude}
In the previous section, we have calculated the tree level $K^+N$ amplitude based on
chiral perturbation theory up to the next-to-leading order.
In this subsection, we carry out the $\chi^{2}$ fitting to the experimental data 
to determine the next-to-leading order low energy constants
in the tree level amplitude.
We use the experimental data of the 
$K^+p$ elastic differential cross sections and $I=0$ total cross section.

We use the isospin averaged masses of nucleon, kaon, $\Lambda$, $\Sigma$, $\Xi$, $\pi$ and
the meson decay constants \cite{pdg} as
\begin{eqnarray}
&&M_{N} =  938.9 \ {\rm MeV/c^{2}}, \\
&&M_{K} =  495.6 \ {\rm MeV/c^{2}}, \\
&&M_{\Lambda} = 1115.7 \ {\rm MeV/c^{2}}, \\
&&M_{\Sigma} = 1193.2 \ {\rm MeV/c^{2}}, \\
&&M_{\Xi} = 1318.3 \ {\rm MeV/c^{2}}, \\
&&M_{\pi} = 138.0 \ {\rm MeV/c^{2}}, \\
&&f_{K} = (1.19 \pm 0.01)f_{\pi} = 110 \ {\rm MeV}
\end{eqnarray}
with $f_{\pi}$=92.4 MeV.
The values of the low energy constants $D$ and $F$ in the leading order chiral Lagrangian 
are determined by the fit to the data of the semileptonic hyperon decay
as reported in Ref. \cite{df}:
\begin{eqnarray}
D = 0.80, \enspace F=0.46.
\end{eqnarray}

The low energy constants $b_{D}$ and $b_{F}$ can be determined by the baryon masses 
using Eqs. (\ref{eq:sigma}) to (\ref{eq:lambda}). 
These parameters shift the chiral limit baryon mass $M_{0}$ to their physical masses. 
Using the isospin averaged masses, we have
\begin{eqnarray}
&&b_{D} = \frac{3 \left(M_{\Sigma} - M_{\Lambda} \right)}{16 M_{K}^{2}} \frac{ 1 + \frac{\hat{m}}{{m_s}} }{1- \frac{\hat{m}}{m_s}}
= 6.41 \times 10^{-5}\enspace{\rm MeV^{-1}}, \label{eq:bd} \\
&&b_{F} = - \frac{M_{\Xi} - M_{N}}{8 M_{K}^{2}} \frac{ 1 + \frac{\hat{m}}{{m_s}} }{1- \frac{\hat{m}}{m_s} }
= -2.09 \times 10^{-4}\enspace{\rm MeV^{-1}} \label{eq:bf}
\end{eqnarray}
with 
\begin{eqnarray}
\frac{\hat{m}}{m_s} = \frac{M_{\pi}^{2}}{2 M_{K}^{2} - M_{\pi}^{2}} = 0.040.
\end{eqnarray}
The rest of the low energy constants are determined 
so as to reproduce the observed cross section by using the $\chi^{2}$ fit. 
The value of $\chi^{2}$ is given by
\begin{eqnarray}
\chi^{2}=\sum_{i}^{N}\left(\frac{y_{i}-f(x_{i})}{\sigma_{i}} \right)^{2}
\label{eq:chi}
\end{eqnarray}
where $y_{i}$,  $f(x_{i})$, $\sigma_{i}$ and $N$ are the experimental data, 
the theoretical calculations which include the parameters, 
the errors of the data and the number of the data, respectively.
We use the partial wave series up to the $G$-wave ($l=4$), 
and we have confirmed the convergence of the partial wave expansion. 
In the isospin basis, we define 
\begin{eqnarray}
  b^{I=1} &=& b_{0} + b_{D},\label{eq:b1}\\
  d^{I=1} &=& 2d_{2} + d_{3} + d_{4}, \\
  b^{I=0} &=& b_{0} - b_{F}, \label{eq:b0}\\
  d^{I=0} &=& 2d_{1} - 2d_{4} + d_{3}. 
\end{eqnarray}

\begin{table}[]
\begin{center}
\caption{Determined parameters for the tree level amplitude.
$b^{I=0}$ can be fixed by $b^{I=1}, b_{D}$ and $b_{F}$.}
\begin{tabular}{|  c |  c  | c | } 
\hline
$b^{I=1}$           &$1.01 \times 10^{-5}$ \enspace ${\rm MeV^{-1}}$           &$\chi^{2}/N = 0.83$                                             \\
$d^{I=1}$           &$4.33 \times 10^{-4}$ \enspace ${\rm MeV^{-1}}$           &      \\
\hline \hline
$b^{I=0}$           &$1.55 \times 10^{-4}$ \enspace ${\rm MeV^{-1}}$           &                                          \\
\hline
$d^{I=0}$           &$3.91 \times 10^{-4}$ \enspace ${\rm MeV^{-1}}$           &$\chi^{2}/N = 12.0$            \\
\hline
\end{tabular} 
\label{tab:para}
\end{center}
\end{table}

First, to determine the low energy constants for $I=1$, $b^{I=1}, d^{I=1}$,
we carry out the $\chi^{2}$ fit for the $K^+p$ elastic differential cross section
using the data at the laboratory momentum $p_{{\rm lab}}$ = 205 MeV/c ($N$=21) \cite{cameron}
where chiral perturbation theory may be applicable.
The best fit of the parameters for $K^+p$ is summarized in the
Table \ref{tab:para}.
At this time, we can determine $b_0$ from $b^{I=1}$ with Eq. (\ref{eq:b1}) as
$b_{0} = -5.40 \times 10^{-5}\ {\rm MeV^{-1}}$.
Using the values of $b_{0}$ and $b_{F}$, the parameter $b^{I=0}$ is calculated 
by Eq. (\ref{eq:b0}) in Table \ref{tab:para}.
Then, we carry out the $\chi^{2}$ fit for $I=0$ total cross section
to determine only $d^{I=0}$
using the data between $p_{{\rm lab}}=$ 366 to 799 MeV/c ($N$=27) \cite{bowen1970, bowen1973, carroll1973}.
The best fit of the parameters for $I=0$ is found in Table \ref{tab:para}.
One should not compare these low energy constants with those determined by fitting the unitarized amplitude to experiments, 
such as the low energy constants obtained in Ref. \cite{Ikeda:2011pi} for $S=-1$ sector. 
As discussed in Ref. \cite{Hyodo:2008xr}, once unitarization is performed, 
the loop contribution requires us to renormalize the tree level amplitude. 
In the renormalization procedure, once the parameters are determined from the experimental values, 
the contributions beyond perturbation are implicitly included in the low energy constants, 
and also the low energy constants get dependent on the renormalization scheme. 
Thus, the low energy constants are not common in different channels any more. 

\begin{figure}[]
 \begin{center}
  \includegraphics[width=100mm]{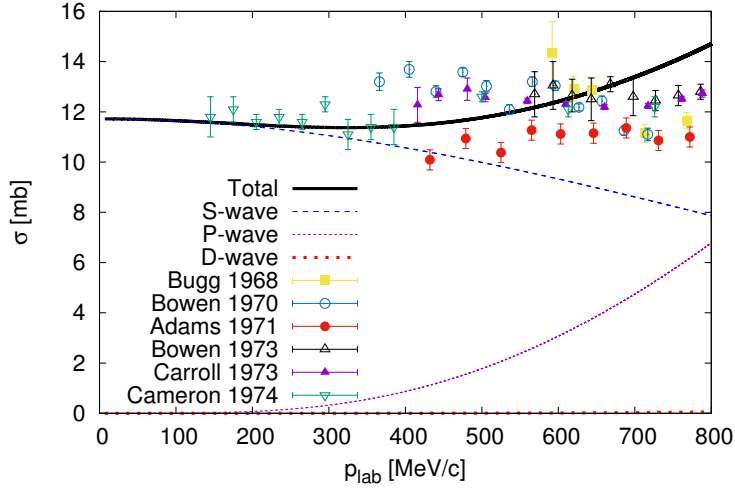}
 \vspace{-0.5cm}
 \caption{The $I=1$ total cross section from chiral perturbation theory up to the next-to-leading order comparison
with the experimental data 
\cite{cameron, bowen1970, bowen1973, carroll1973, bugg1968, adams1971}.
The partial wave contributions are plotted in the dashed lines.
The horizontal axis means the $K^{+}$ meson incident momentum in the lab frame $p_{{\rm lab}}$ in the unit of MeV/c
and the vertical axis means the total cross section $\sigma$ in the unit of mb.}
 \label{fig:tot1_Tree}
 \end{center}
\end{figure}

\begin{figure}[]
 \begin{center}
  \includegraphics[width=100mm]{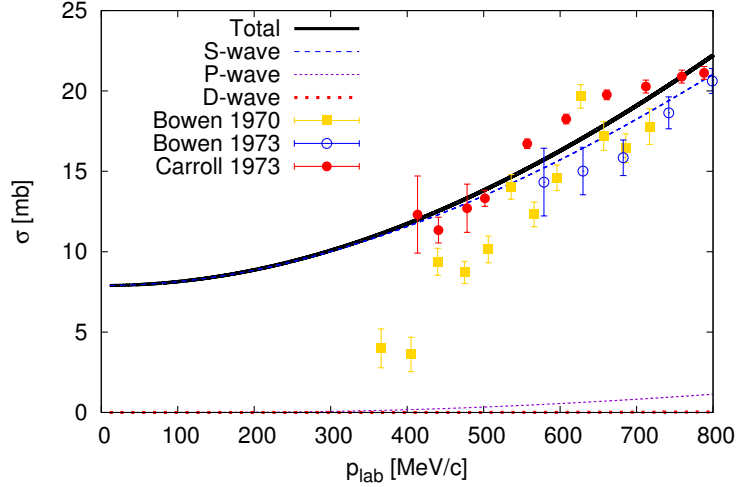}
 \end{center}
  \vspace{-0.5cm}
 \caption{The $I=0$ total cross section from chiral perturbation theory up to the next-to-leading order 
 comparison with the experimental data \cite{bowen1970, bowen1973, carroll1973}.}
  \label{fig:tot0_Tree}
\end{figure}

\begin{figure}[]
\begin{minipage}{0.5\hsize}
  \begin{center}
   \includegraphics[width=80mm]{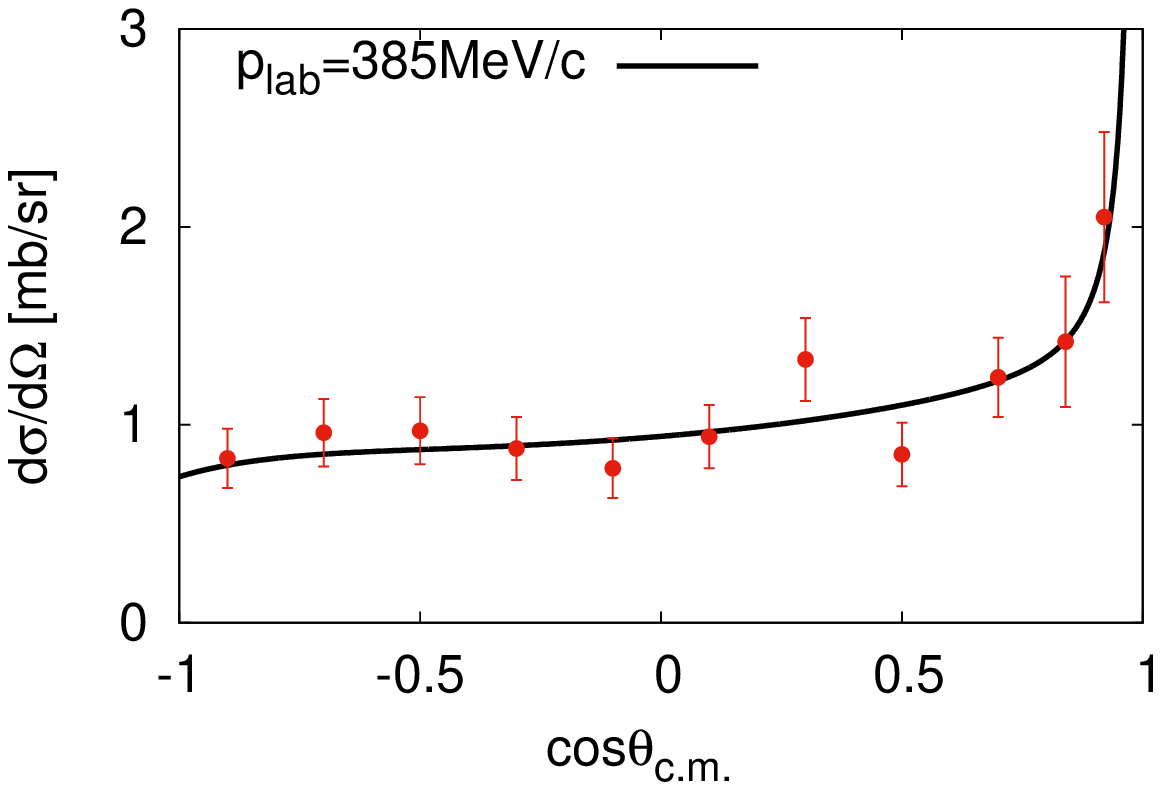}
  \end{center}
  \end{minipage}
\begin{minipage}{0.5\hsize}
  \begin{center}
   \includegraphics[width=80mm]{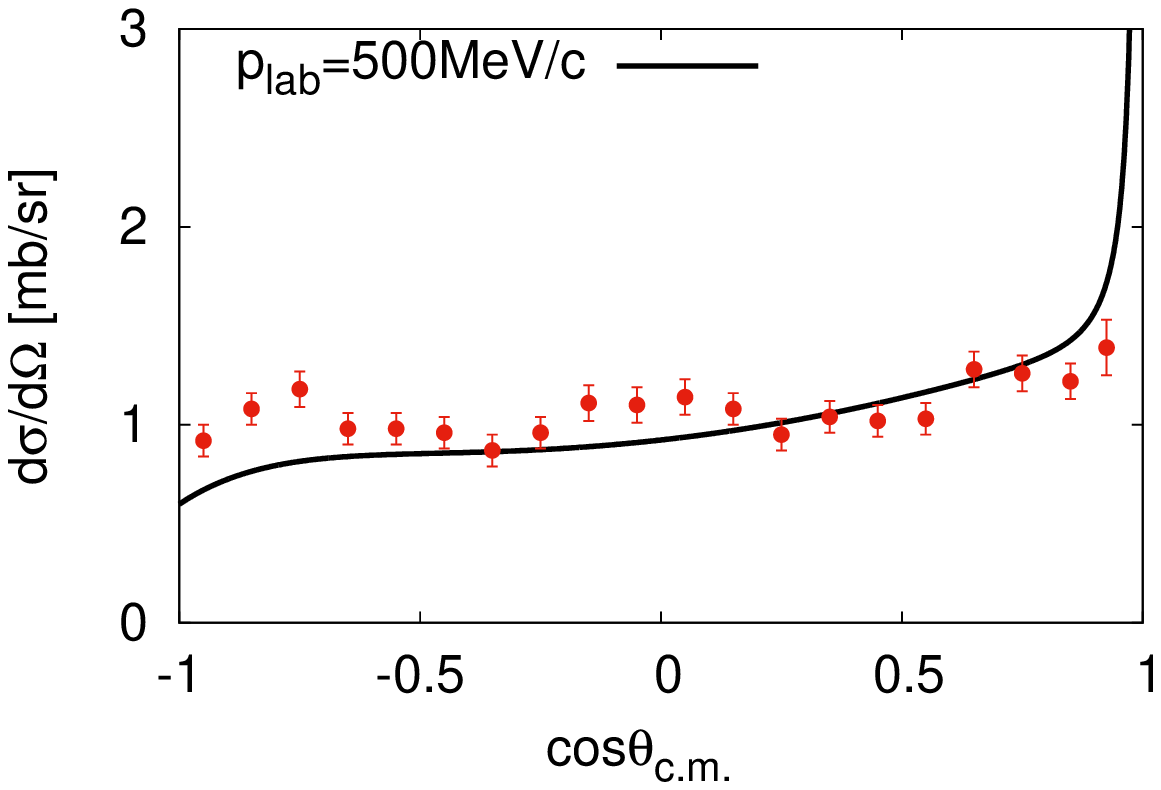}
  \end{center}
   \end{minipage}
  \vspace{-0.5cm}
\caption{
The differential cross section of $K^{+}p$ 
elastic scattering at several lab momentum $p_{{\rm lab}}$
compared with the experimental data of Ref. \cite{cameron}.
The differential cross section $d\sigma/d\Omega$ is shown  
in units of mb/sr as a function of cos$\theta_{{\rm c.m.}}$.}
\label{fig:kp_Tree_diff}
\end{figure}

\begin{figure}[]
 \begin{minipage}{0.5\hsize}
  \begin{center}
   \includegraphics[width=80mm]{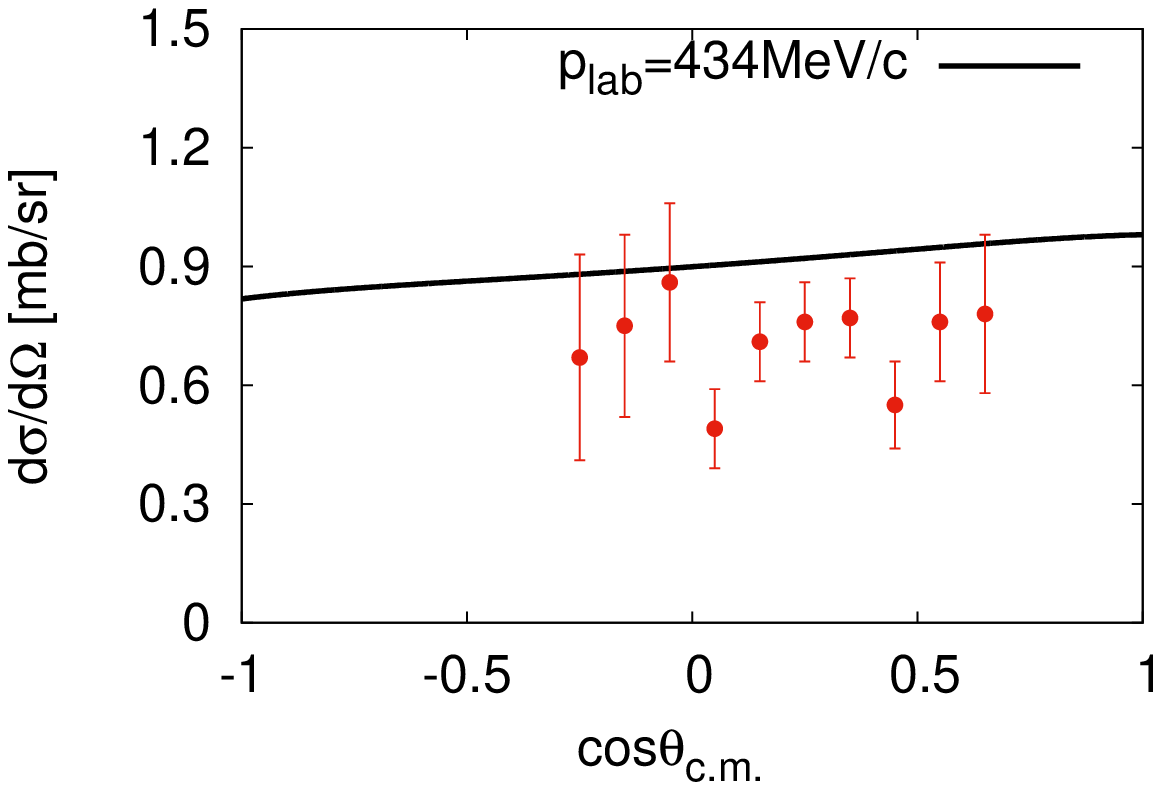}
  \end{center}
\end{minipage}
 \begin{minipage}{0.5\hsize}
  \begin{center}
   \includegraphics[width=80mm]{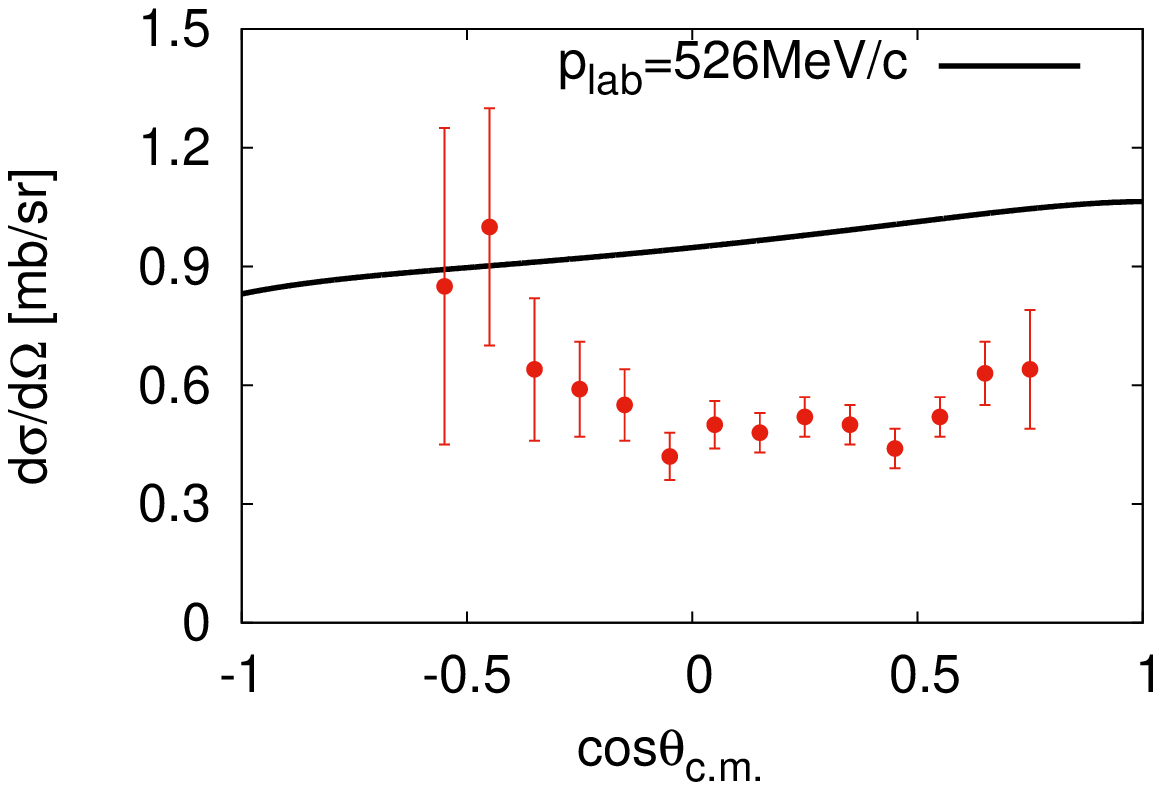}
  \end{center}
 \end{minipage}
 \vspace{-0.5cm}
 \caption{The differential cross section of $K^{+}n$ elastic scattering 
 compared with the experimental data of Ref. \cite{dam1975}.}
\label{fig:kn_Tree_diff}
 \end{figure}

In the following, we compare our results with the experimental values for
energies up to 800 MeV/c, from which inelastic contributions
like pion production start to be significant. 
In Figs. \ref{fig:tot1_Tree} and \ref{fig:tot0_Tree}, 
we show the results of the total cross sections calculated with the tree level amplitude, 
obtained by chiral perturbation theory up to the next-to-leading order.
In spite of the perturbative calculation, we find substantially nice agreement with the data 
up to $p_{\rm lab} =600$ MeV/c. 
It is also found that the contribution of the partial wave series up to the $P$-wave is dominant.
In Fig. \ref{fig:kp_Tree_diff}, 
we show the differential cross section of the $K^{+}p$ elastic scattering together with
the experimental data \cite{cameron}. 
In the theoretical calculation, we include the Coulomb correction 
discussed in Sec. \ref{sec:coulomb}.
Although we have used only the data at $p_{\rm lab}=205$ MeV/c 
to determine the parameters, surprisingly we obtain great reproduction of the differential cross 
sections for lower and higher energies up to $p_{{\rm lab}}$ = 500 MeV/c. 
In Fig. \ref{fig:kn_Tree_diff}, we show the differential cross section of the $K^{+}n$ 
elastic scattering at $p_{{\rm lab}}$ = 434 and 526 MeV/c
with the experimental data taken from Ref. \cite{dam1975}.
Although we reproduce well $I=0$ total cross section in the low energy region
with the parameter determined by $I=0$ total cross section 
as shown in Fig. \ref{fig:tot0_Tree}, 
the reproduction of the $K^{+}n$ differential cross section 
at $p_{{\rm lab}}=$526 MeV/c is not so impressive. 


\begin{figure}[]
 \begin{minipage}{0.5\hsize}
  \begin{center}
   \includegraphics[width=80mm]{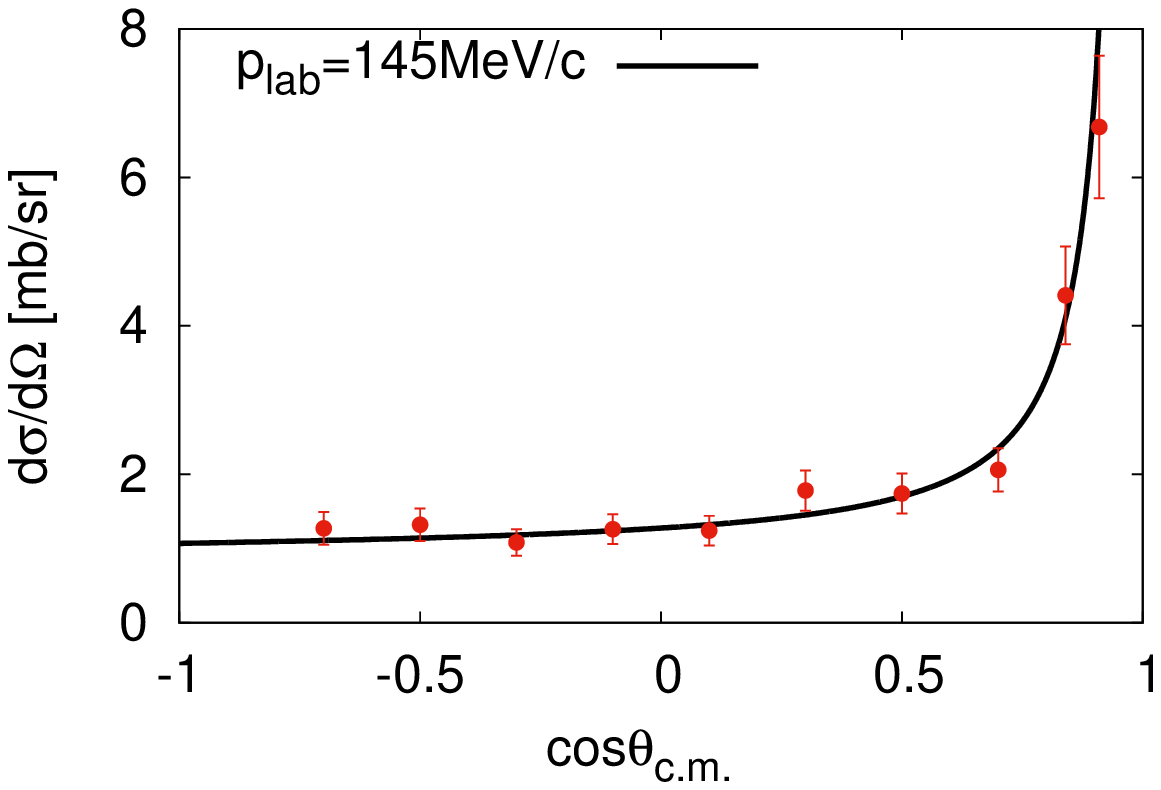}
  \end{center}
 \end{minipage}
 \begin{minipage}{0.5\hsize}
  \begin{center}
   \includegraphics[width=80mm]{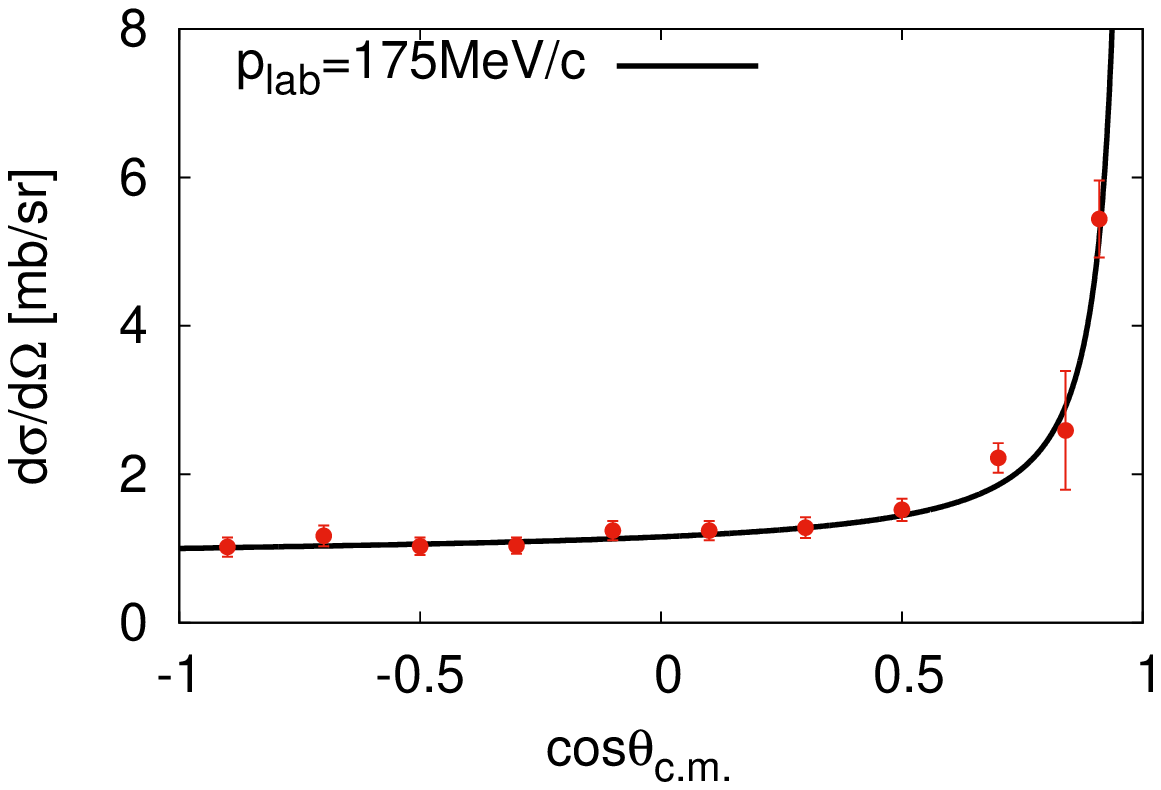}
  \end{center}
 \end{minipage}
\begin{minipage}{0.5\hsize}
  \begin{center}
   \includegraphics[width=80mm]{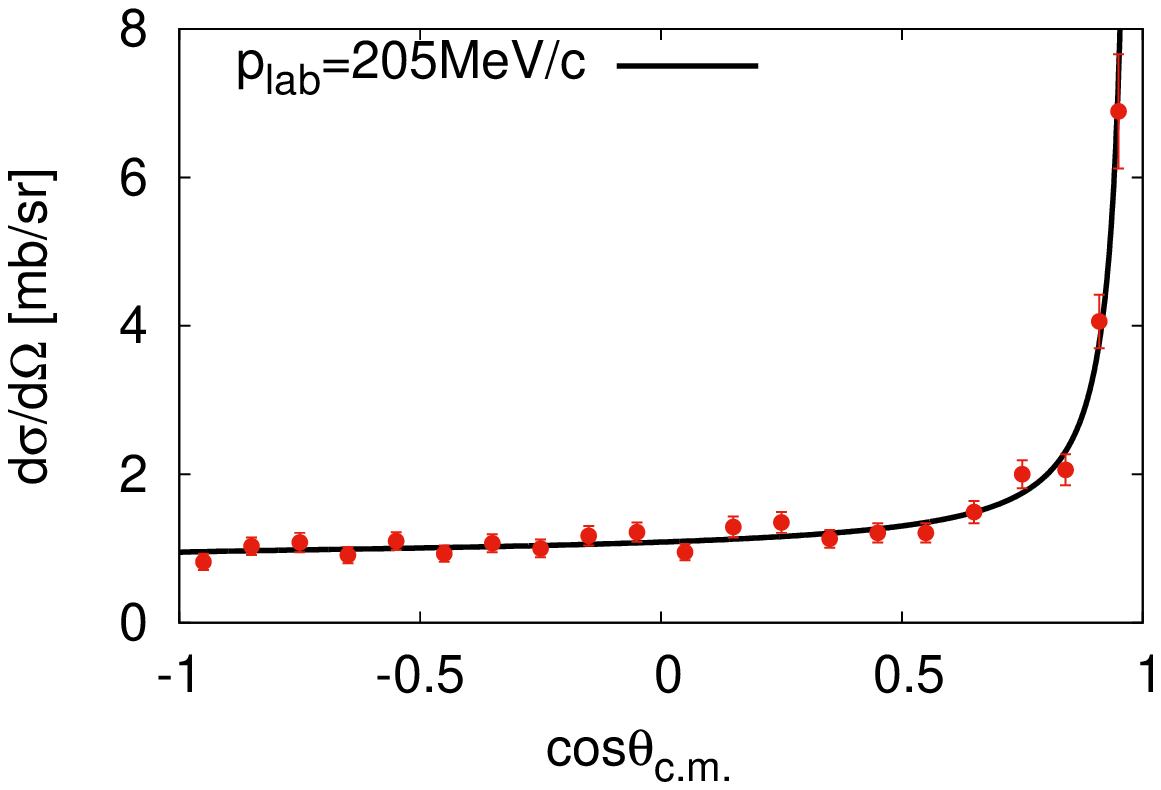}
  \end{center}
 \end{minipage}
\begin{minipage}{0.5\hsize}
  \begin{center}
   \includegraphics[width=80mm]{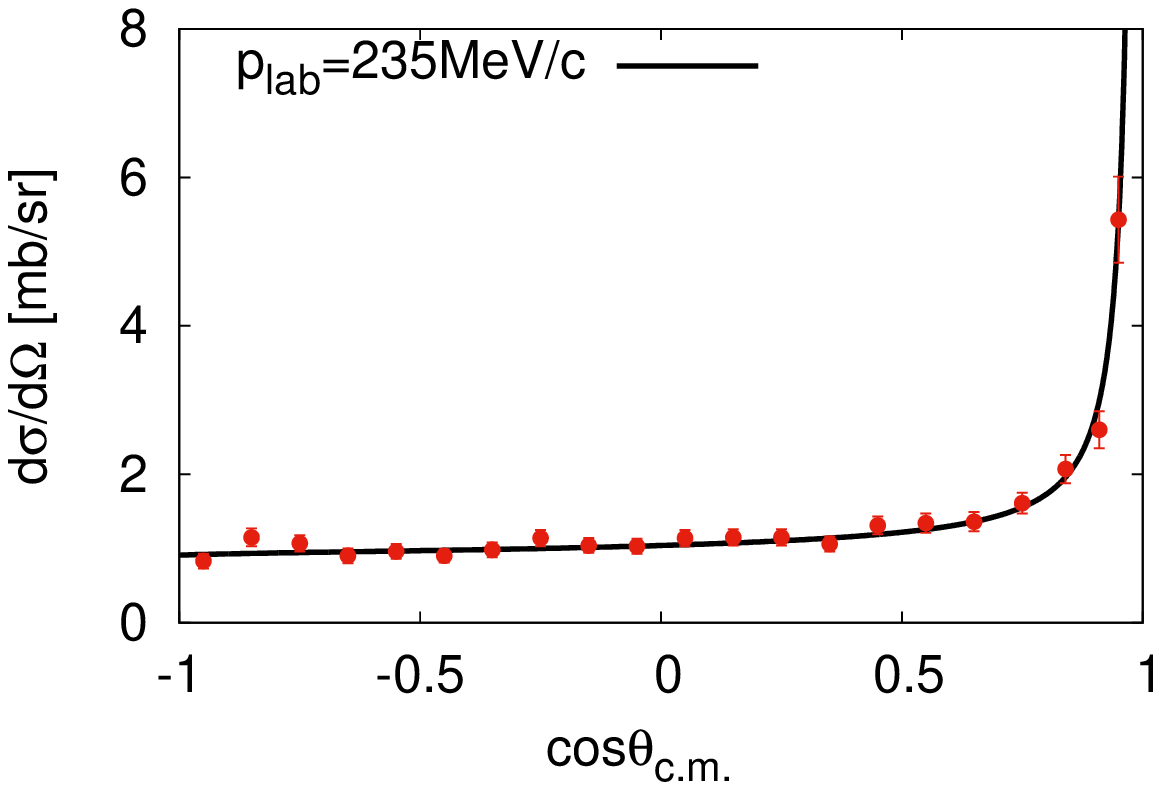}
  \end{center}
  \end{minipage}
 \begin{minipage}{0.5\hsize}
  \begin{center}
   \includegraphics[width=80mm]{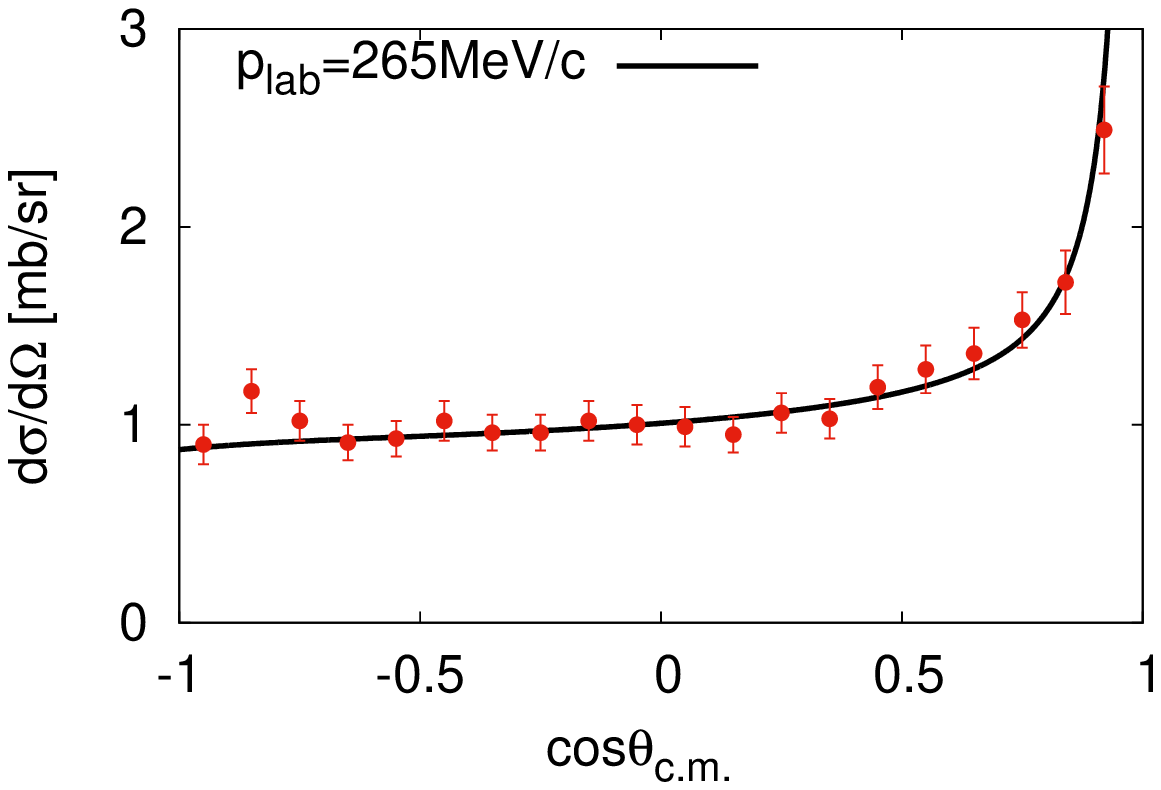}
  \end{center}
 \end{minipage}
\begin{minipage}{0.5\hsize}
  \begin{center}
   \includegraphics[width=80mm]{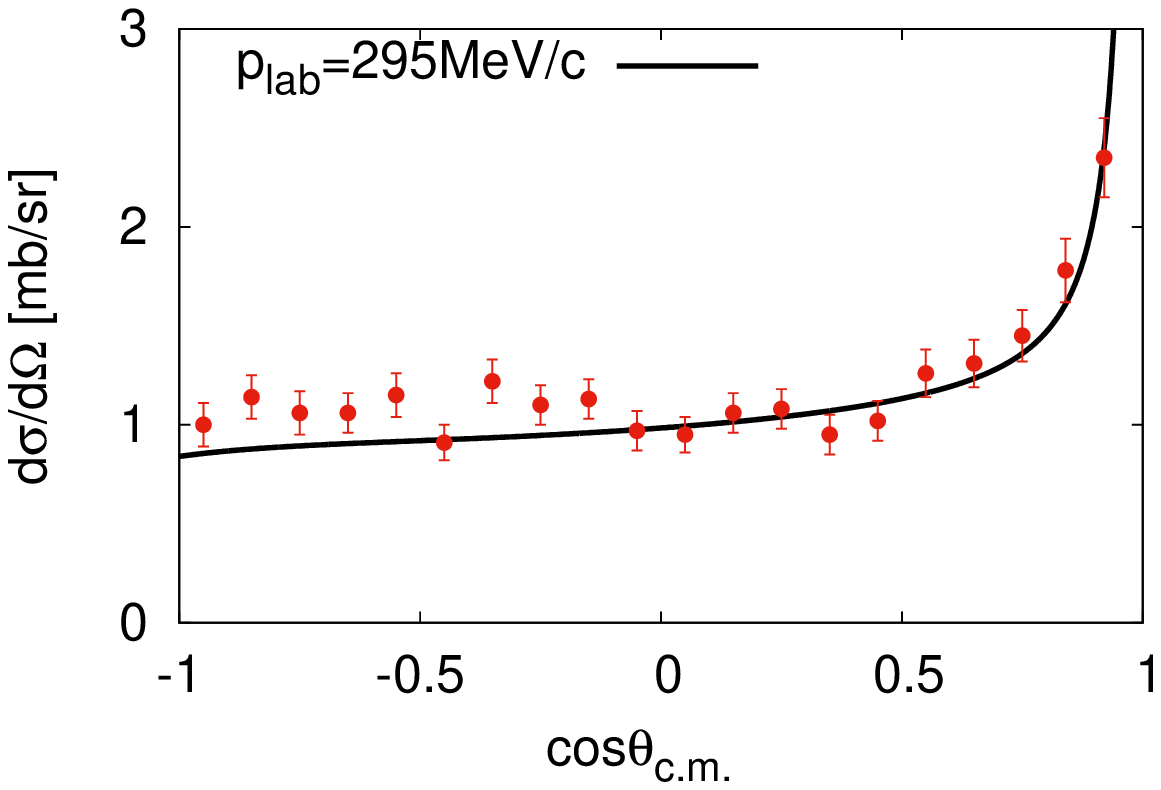}
  \end{center}
 \end{minipage}
\begin{minipage}{0.5\hsize}
  \begin{center}
   \includegraphics[width=80mm]{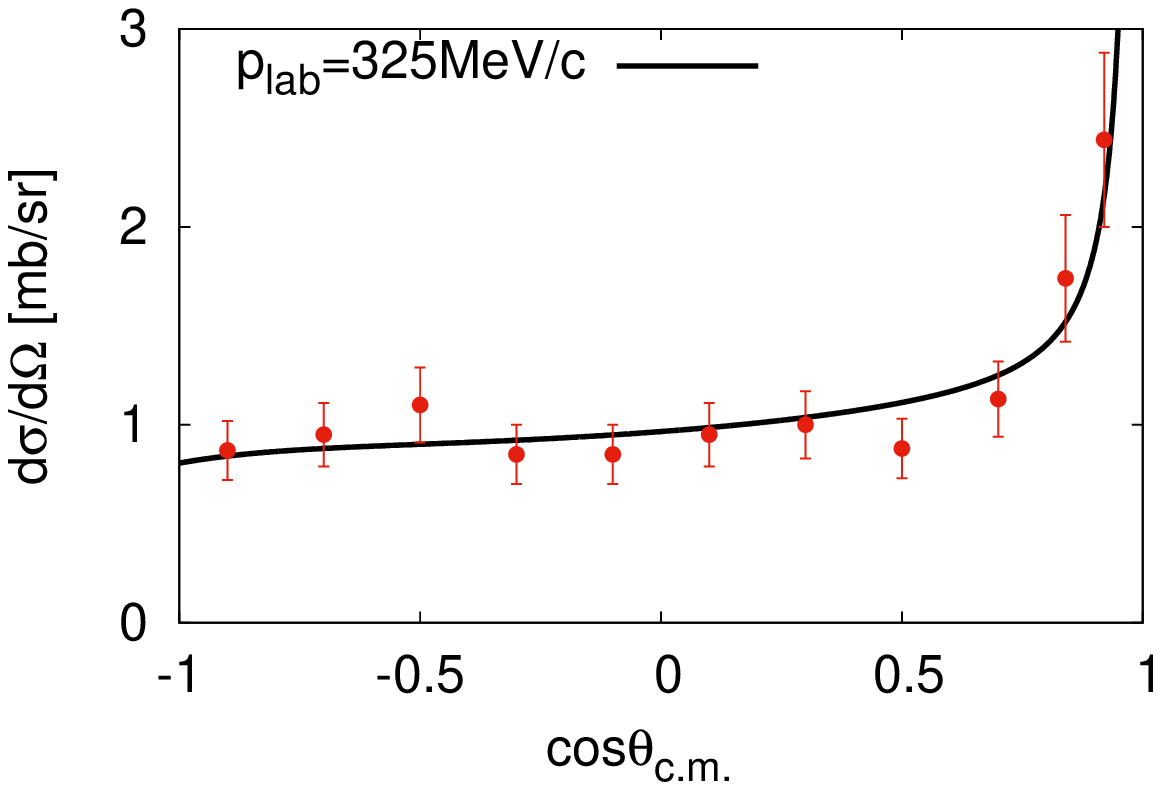}
  \end{center}
   \end{minipage}
\begin{minipage}{0.5\hsize}
  \begin{center}
   \includegraphics[width=80mm]{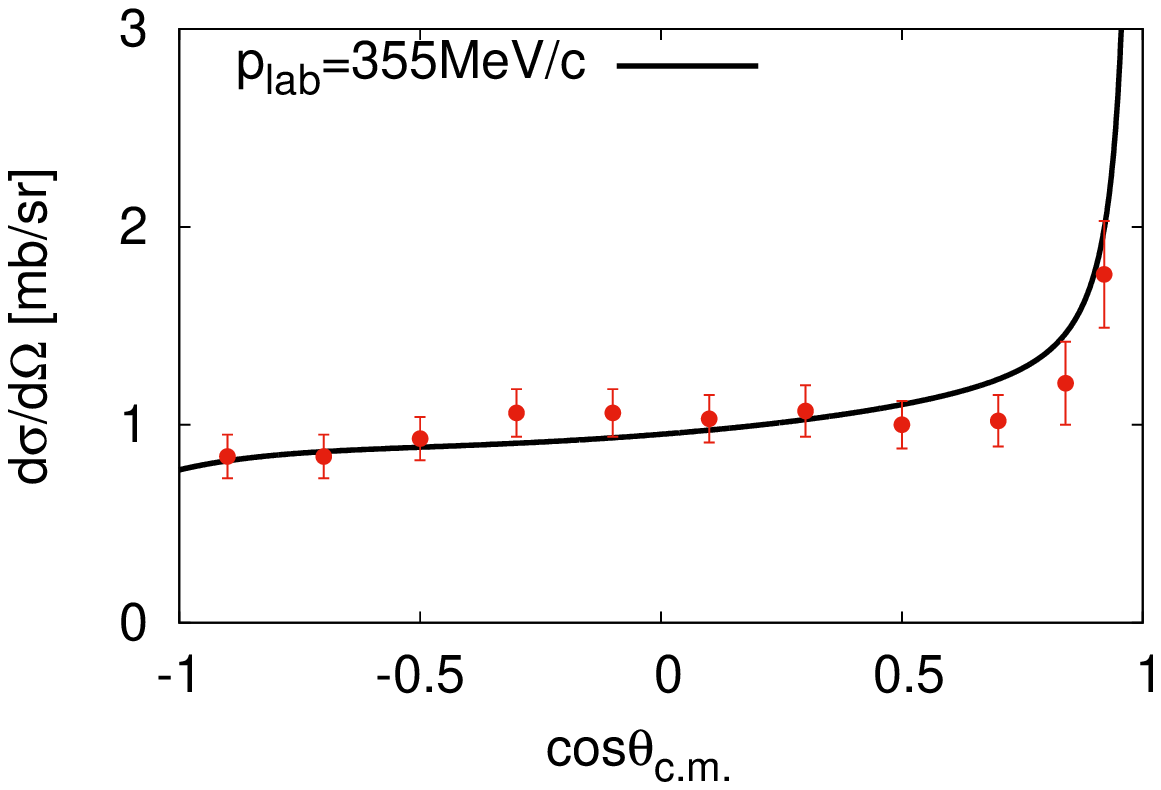}
  \end{center}
  \end{minipage}
 \end{figure}

\subsection{Wavefunction renormalization}
We have obtained a good description of the tree level $K^{+}N$ amplitude in the low energies.   
In this subsection, we calculate
the wavefunction renormalization using the amplitude calculated above by following the method
summarized in Sec. \ref{sec:2}. 
The self-energy of $K^{+}$ in the nuclear medium is calculated by Eq.~(\ref{eq:selfene})
with the forward $KN$ elastic amplitude by taking account of the Fermi motion of the nucleons. 
We assume symmetric nuclear matter and evaluate the wavefunction renormalization at the 
saturation density $\rho_{0}$. For symmetric nuclear matter, we consider the averaged amplitude 
of the $K^{+}p$ and $K^{+}n$ elastic scattering, which is given in terms of the isospin amplitudes by
\begin{equation}
 T(\theta_{\rm c.m.} = 0) = \frac{1}{4} [ 3 T^{I=1}(\theta_{\rm c.m.} = 0) 
 + T^{I=0}(\theta_{\rm c.m.} = 0)].
\end{equation}

Before going into the numerical calculation, 
we show a simple analytic calculation, which is good to illustrate the method of the wavefunction renormalization \cite{j2016}.
We consider the model-independent Weinberg-Tomozawa term 
at the tree level which is leading order chiral perturbation theory.
The Weinberg-Tomozawa term is given by 
\begin{eqnarray}
T^{I=0}_{WT}=0,\qquad T^{I=1}_{WT}=\frac{1}{2f_{K}^{2}}\bar{u}(\vec{p}_{4}, s) \left(\slash{p_{1}} +\slash{p_{3}}\right) u(\vec{p}_{2}, s)
\end{eqnarray}
as we saw in Sec. \ref{sec:3}.
The isospin averaged tree level forward amplitude for Weinberg-Tomozawa term is given by
\begin{eqnarray}
T (\theta_{{\rm c.m.}} = 0) = \frac{3}{4} \frac{\omega^{\ast}}{f_{K}^{2}}.
\label{eq:wt}
\end{eqnarray}
Using Eq. (\ref{eq:wt}), we calculate the wavefunction renormalization factor $Z$ 
at the threshold
$\omega^{\ast} = M_{K}$ without Fermi motion using Eqs. (\ref{eq:Trho}) and (\ref{eq:z_factor})
\begin{align}
Z &= 1 +\frac{3 \rho_{0}}{8 M_{K} f_{K}^{2}}  \frac{\rho}{\rho_{0}} \nonumber \\
   &= 1+ 0.082 \frac{\rho}{\rho_0}.
\end{align}
This result means that the magnitude of the amplitude is increased by about 8\% 
at the saturation density due to the wavefunction renormalization.

\begin{figure}[]
 \begin{center}
  \includegraphics[width=100mm]{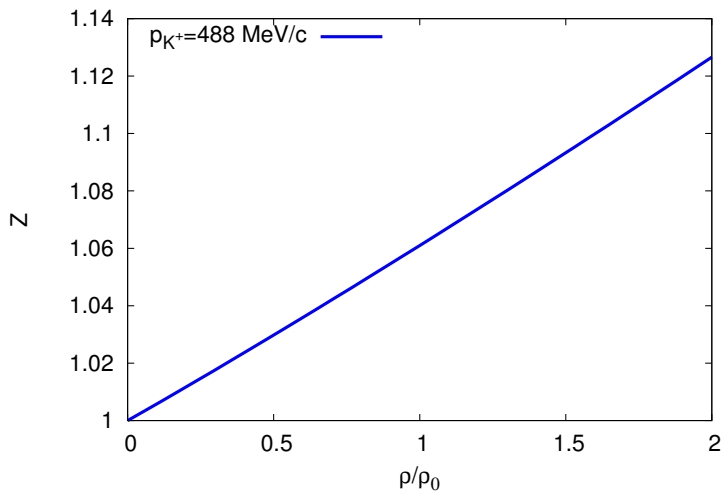}
 \vspace{-0.5cm}
 \caption{The density dependence of the wavefunction renormalization factor $Z$ with Fermi motion
 at the $p_{{K^{+}}} = 488$ MeV/c using the tree level amplitude.}
 \label{fig:rho488}
 \end{center}
\end{figure}

Now let us show the wavefunction renormalization factor obtained from the $K^{+}N$ scattering calculated 
by chiral perturbation theory up to the next-to-leading order.
The results for the wavefunction renormalization factor 
are summarized in Table \ref{tab:wf_tree}. 
In the middle columns, 
we show the results without Fermi motion of the nucleons which is calculated by Eq. (\ref{eq:Trho}), 
while in the right columns we have the results with the Fermi motion calculated 
by Eq. (\ref{eq:selfene}). 
The table shows that the wavefunction renormalization factor gives 2 to 6\% enhancement 
for the in-medium self-energy at the saturation density. 
This shows that the wavefunction renormalization could explain a part of 
the breakdown of the linear density approximation for the $K^{+}$-nucleus scattering. 
We also find in Table \ref{tab:wf_tree} that the effect of 
the Fermi motion for the nucleons is minor. 
In Fig. \ref{fig:rho488}, we show also the density dependence of 
the wavefunction renormalization factor with the Fermi motion at the $p_{K^{+}}=488$ MeV/c. 
Because of the minor contribution of the Fermi motion, 
the density dependence is almost linear in this approximation. 
For further detailed study of the medium effects on $K^{+}$, 
we need to develop in-medium chiral perturbation theory 
for the $K$ meson similar for the pion developed, for example, 
in Refs. \cite{Goda:2013bka, Goda:2013npa}.

\begin{table}[]
\begin{center}
\caption{The wavefunction renormalization factor without the Fermi motion $Z_{{\rm w/o}}$ and with the Fermi motion $Z$
obtained by chiral perturbation theory up to the next-to-leading order.
$p_{K^{+}}$ is the $K^{+}$ momentum at the nuclear matter rest frame.}
\begin{tabular}{| c ||  c | c |  } 
\hline
$p_{K^{+}}$ [MeV/c]&  $Z_{{\rm w/o}}$&         $Z$   \\ \hline \hline
0.0    &  1.05&        1.06              \\
130.0&  1.05&        1.05               \\
488.0&  1.03&       1.03      \\    
531.0&  1.03&       1.03         \\
656.0&  1.02&      1.02          \\
714.0&  1.02&      1.02        \\
\hline
\end{tabular} 
\label{tab:wf_tree}
\end{center}
\end{table}

\subsection{Unitarized amplitude and possible resonance in $I=0$}
We also see the unitarized amplitude of the tree level amplitude
by following the method developed for the $\bar{K}N$ scattering where the $\Lambda(1405)$
is dynamically generated \cite{lambda, ksw1995, or1998, orb2002}.
The way to perform the unitarization is summarized in appendix \ref{sec:appendix}.

First, to determine the the low energy constants for $I=1$, $b^{I=1}, d^{I=1}$
and the subtraction constant $a^{I=1}$,
we carry out the $\chi^{2}$ fit for $K^+p$ elastic differential cross section
using the data at the $p_{{\rm lab}}$ = 726 MeV/c ($N$=20) \cite{cameron}
where $P$-wave contribution is well constrained.
In the point of view of chiral perturbation theory, 
one may think that the momentum $p_{{\rm lab}}$ = 726 MeV/c 
would be out of applicable energies of chiral perturbation theory,
while carrying out unitarization makes applicable region wider.
If we carry out $\chi^{2}$ fitting at the low energy, such as $p_{{\rm lab}} =$ 205 MeV/c,
the reproduction of the data above $p_{{\rm lab}} =$ 385 MeV/c gets worse.
The best fit of the parameters for $K^+p$ is summarized 
in Table \ref{tab:para2}.
To determine the low energy constants for $I=0$, $b^{I=0}, d^{I=0}$
and the subtraction constant $a^{I=0}$,
we carry out the $\chi^{2}$ fit for $I=0$ total cross section
between $p_{\rm lab} = 336$ to $799$ MeV/c $(N=27)$
using the data \cite{bowen1970, bowen1973, carroll1973}.
Since the relations (\ref{eq:bd}) and (\ref{eq:bf}) are realized in the tree level,
we should treat the low energy constants $b^{I=0}$ and $d^{I=0}$ as free parameters
for the unitarized amplitude.
The best fit of the parameters for $I=0$ is found in Table \ref{tab:para2}.

\begin{table}[]
\begin{center}
\caption{Determined parameters for the unitarized amplitude.}
\begin{tabular}{|  c |  c  | c | } 
\hline
$b^{I=1}$           &$6.18 \times 10^{-4} \enspace {\rm MeV^{-1}}$   &       \\
$d^{I=1}$           &$3.00 \times 10^{-4} \enspace {\rm MeV^{-1}}$  &$\chi^{2}/N = 0.68$       \\
$a^{I=1}$                    &$-1.224$                                                                             &    \\
\hline \hline
$b^{I=0}$           &$4.12 \times 10^{-4} \enspace {\rm MeV^{-1}}$   &       \\
$d^{I=0}$           &$-9.78 \times 10^{-4} \enspace {\rm MeV^{-1}}$  &$\chi^{2}/N = 6.85$       \\
$a^{I=0}$                    &$-1.231$                                                                             &    \\
\hline
\end{tabular} 
\label{tab:para2}
\end{center}
\end{table}

\begin{figure}[]
 \begin{center}
  \includegraphics[width=100mm]{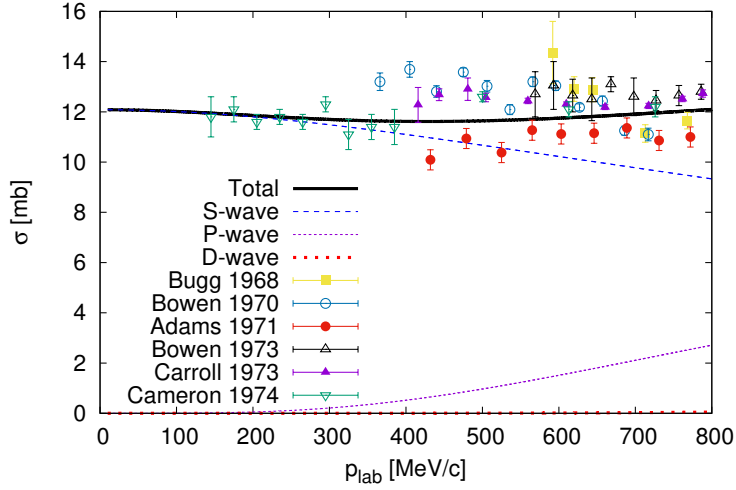}
   \vspace{-0.5cm}
 \caption{The $I=1$ total cross section from unitarization
 comparison with the experimental data \cite{cameron, bowen1970, bowen1973, carroll1973, bugg1968, adams1971}.}
 \label{fig:tot1_Uni}
 \end{center}
\end{figure}

\begin{figure}[]
 \begin{center}
  \includegraphics[width=100mm]{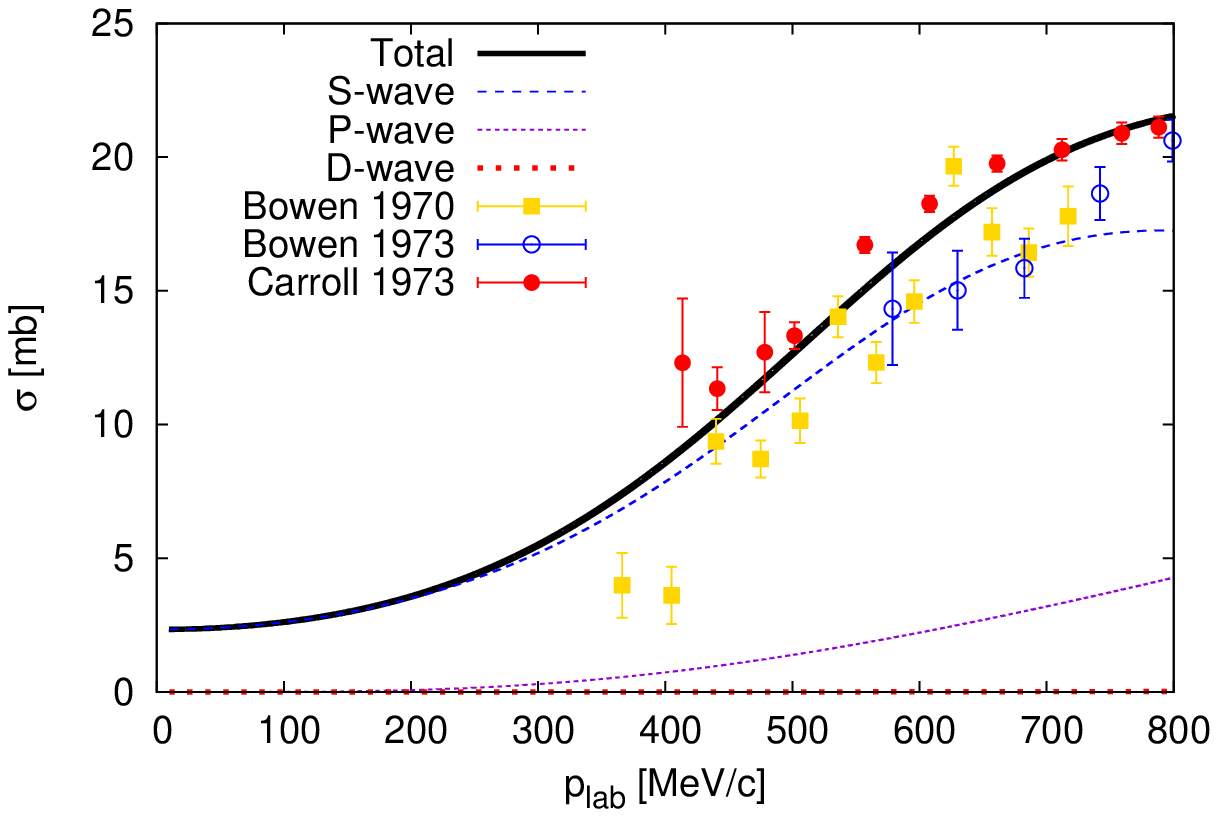}
 \end{center}
  \vspace{-0.5cm}
 \caption{The $I=0$ total cross section from unitarization
  comparison with the experimental data \cite{bowen1970, bowen1973, carroll1973}.}
 \label{fig:tot0_Uni}
\end{figure}

\begin{figure}[]
\begin{minipage}{0.5\hsize}
  \begin{center}
   \includegraphics[width=80mm]{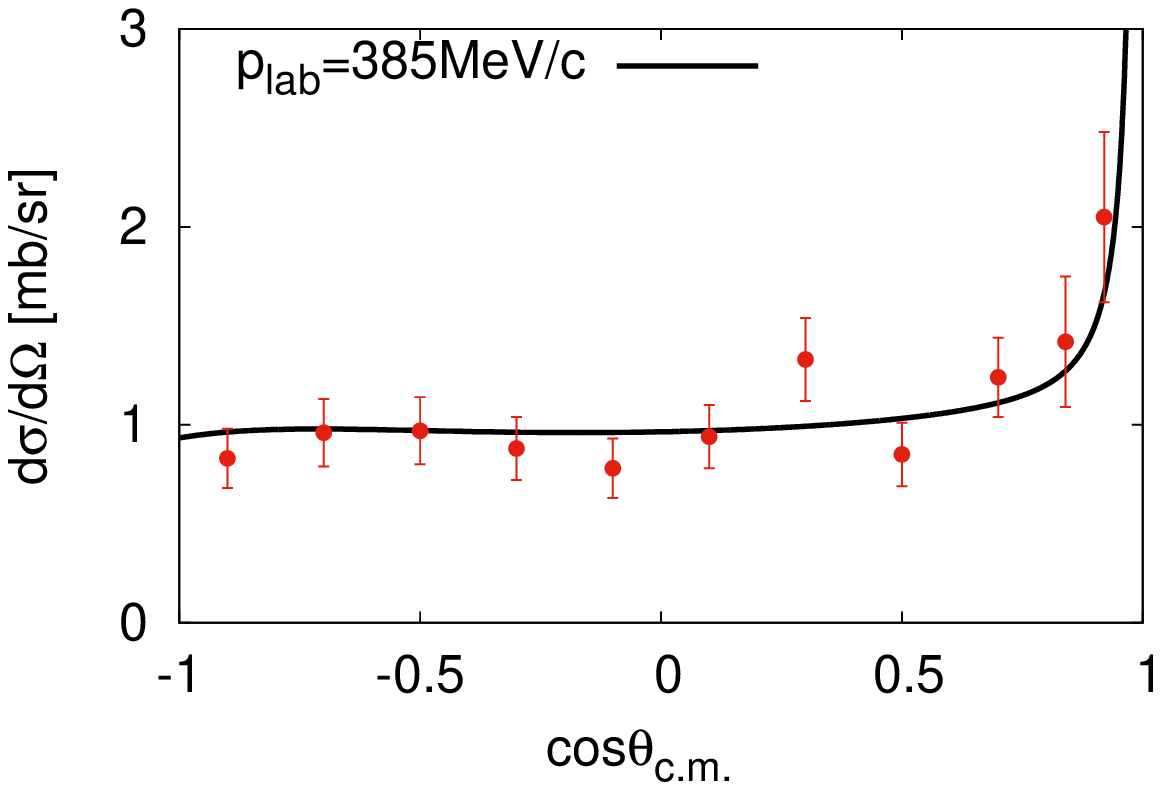}
  \end{center}
 \end{minipage}
\begin{minipage}{0.5\hsize}
  \begin{center}
   \includegraphics[width=80mm]{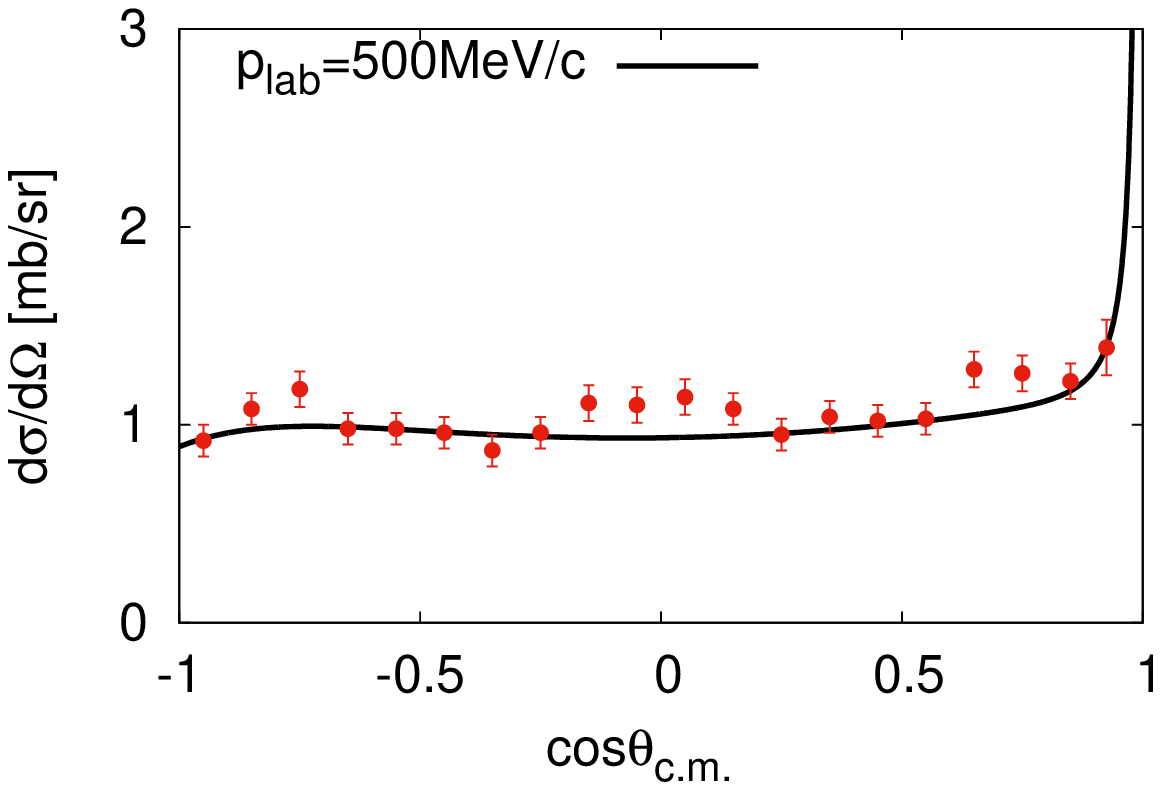}
  \end{center}
  \end{minipage}
\begin{minipage}{0.5\hsize}
  \begin{center}
   \includegraphics[width=80mm]{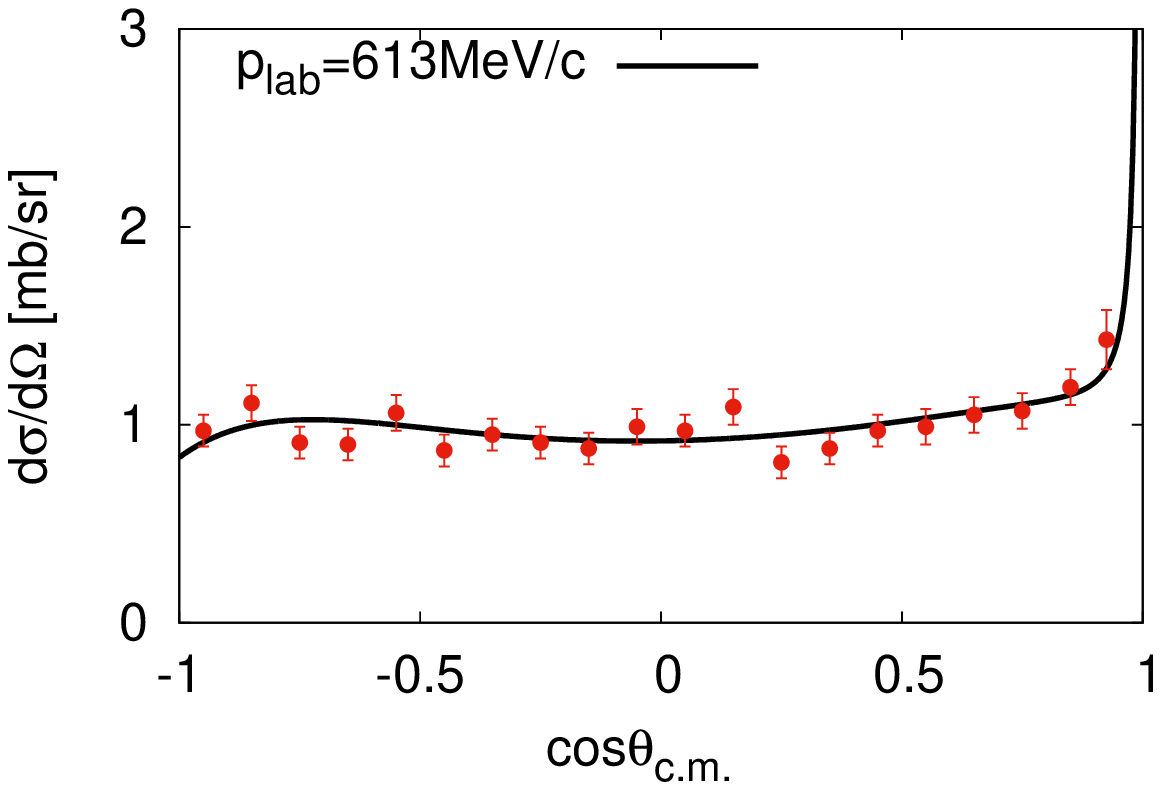}
  \end{center}
  \end{minipage}
\begin{minipage}{0.5\hsize}
  \begin{center}
   \includegraphics[width=80mm]{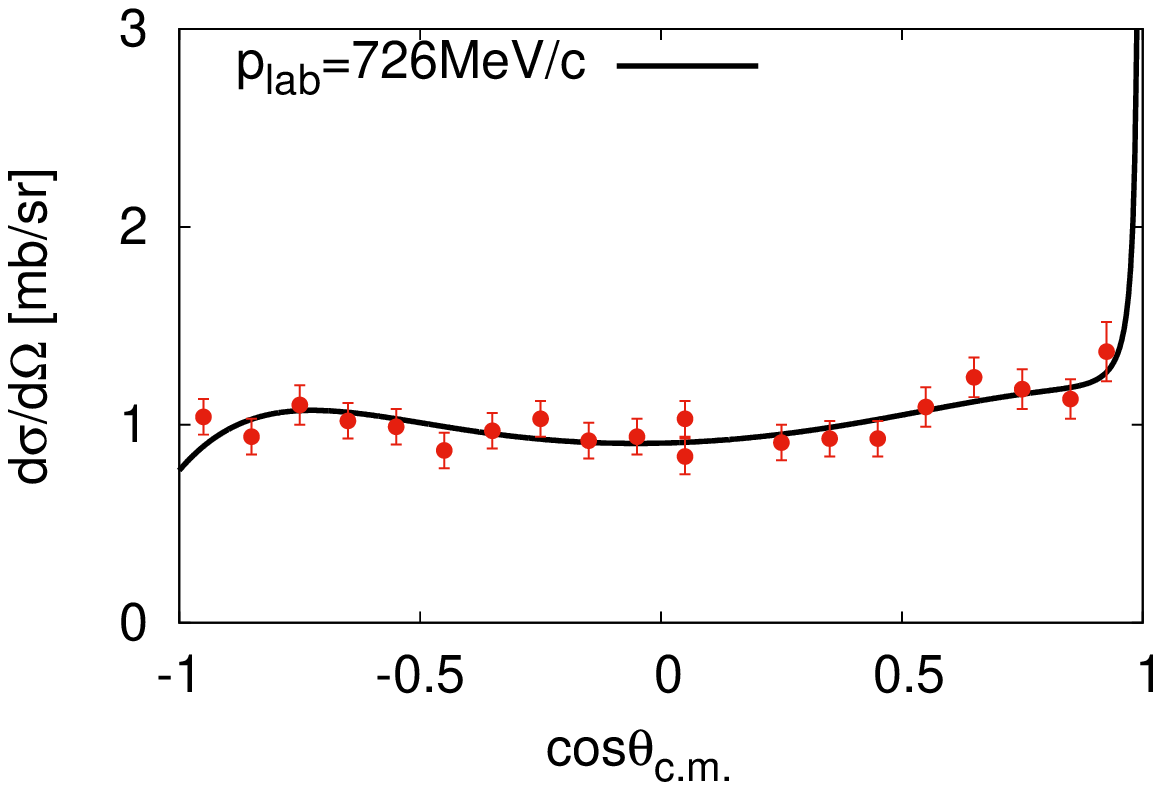}
  \end{center}
 \end{minipage}
 \vspace{-0.5cm}
\caption{The differential cross section of $K^{+}p$
elastic scattering from unitarization at several lab momentum $p_{{\rm lab}}$
compared with the experimental data of Ref. \cite{cameron}.}
\label{fig:kp_diff_uni}
\end{figure}

\begin{figure}[]
 \begin{minipage}{0.5\hsize}
  \begin{center}
   \includegraphics[width=80mm]{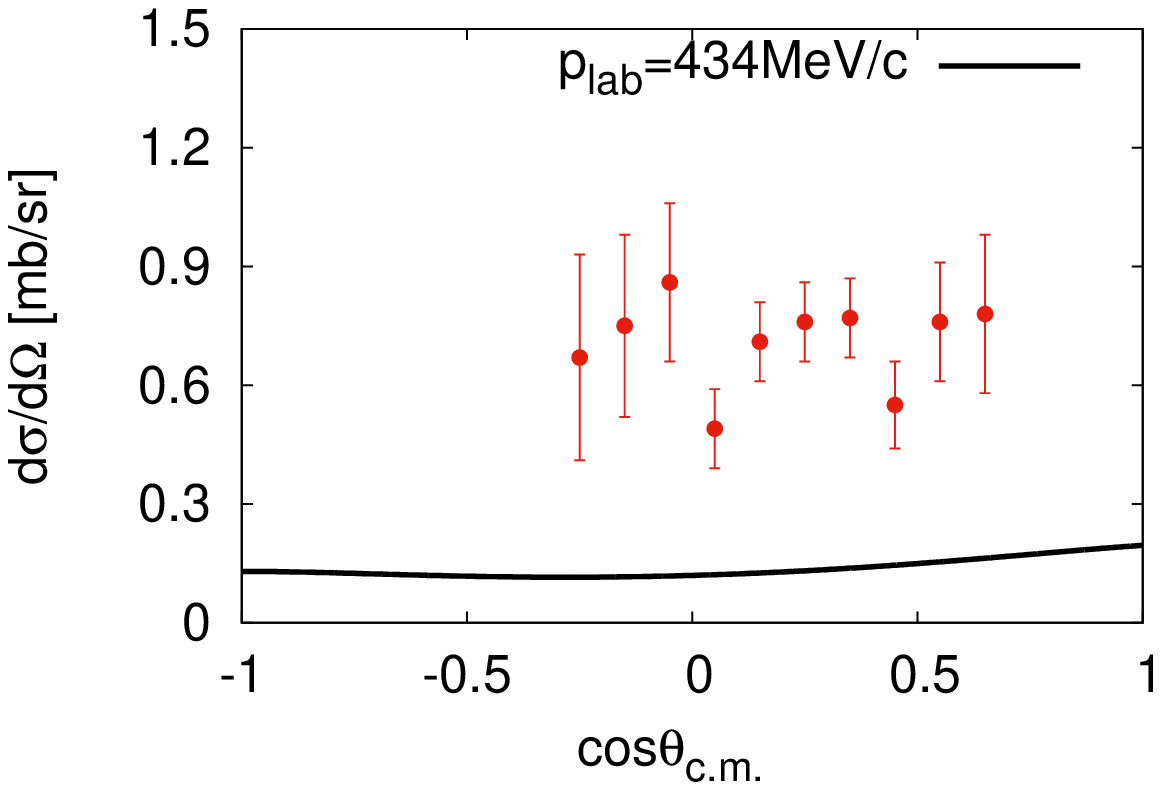}
  \end{center}
 \end{minipage}
 \begin{minipage}{0.5\hsize}
  \begin{center}
   \includegraphics[width=80mm]{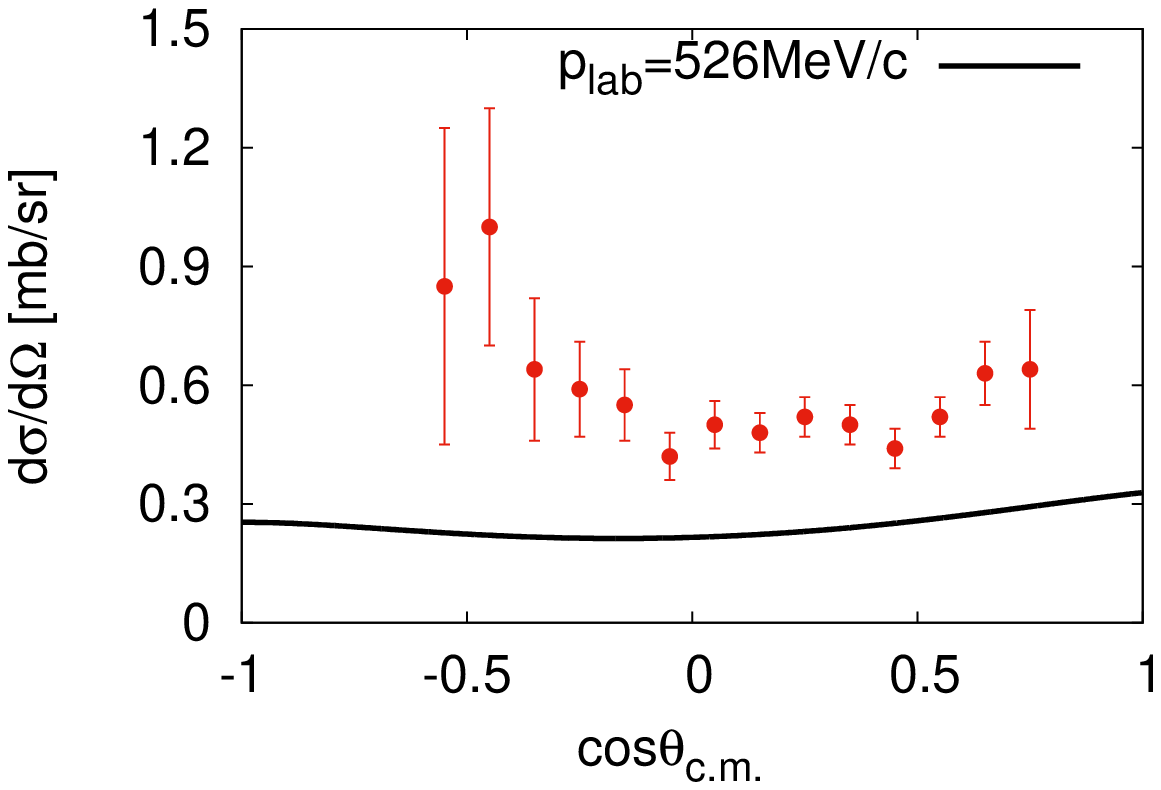}
  \end{center}
  \end{minipage}
\begin{minipage}{0.5\hsize}
  \begin{center}
   \includegraphics[width=80mm]{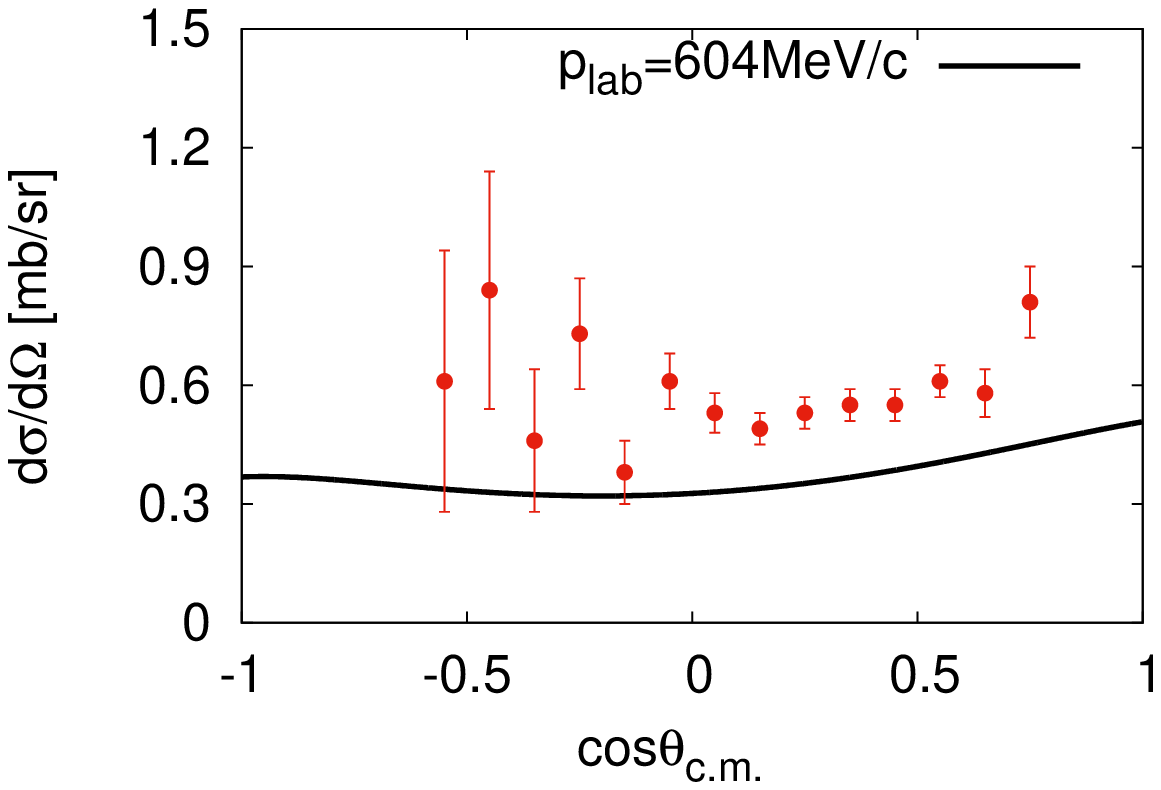}
  \end{center}
 \end{minipage}
\begin{minipage}{0.5\hsize}
  \begin{center}
   \includegraphics[width=80mm]{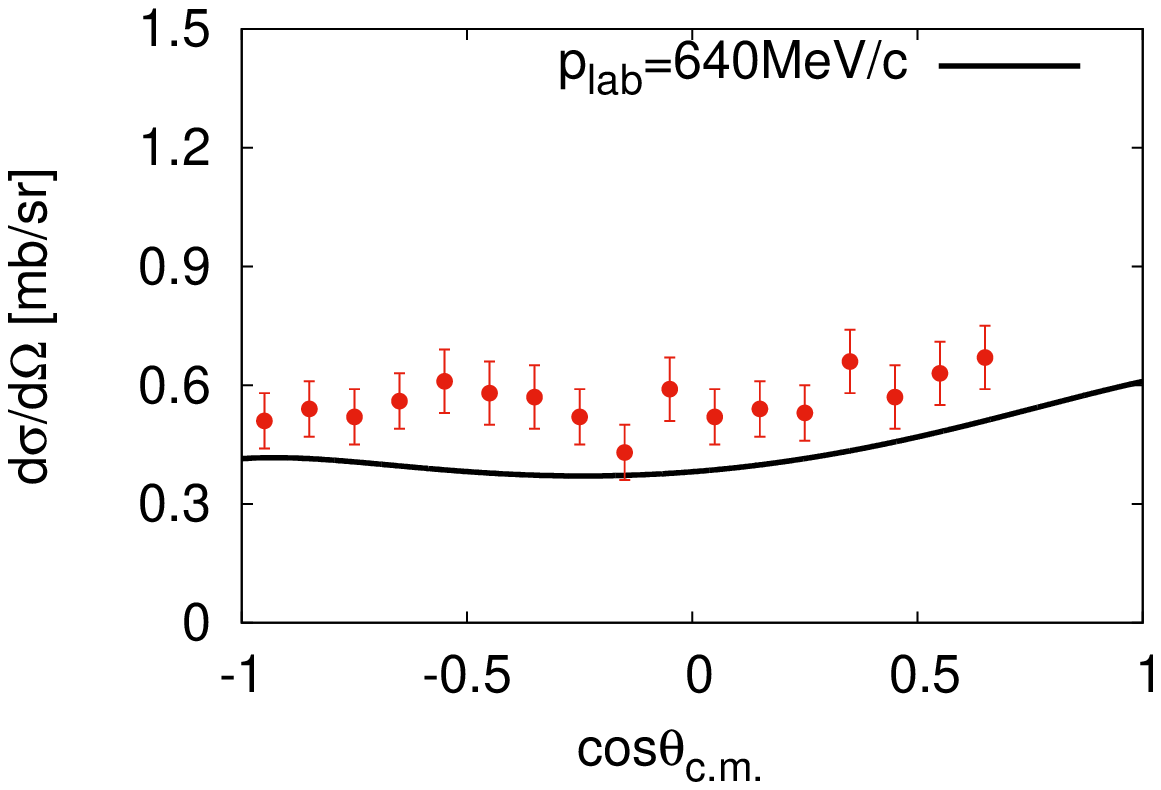}
  \end{center}
 \end{minipage}
 \begin{minipage}{0.5\hsize}
  \begin{center}
   \includegraphics[width=80mm]{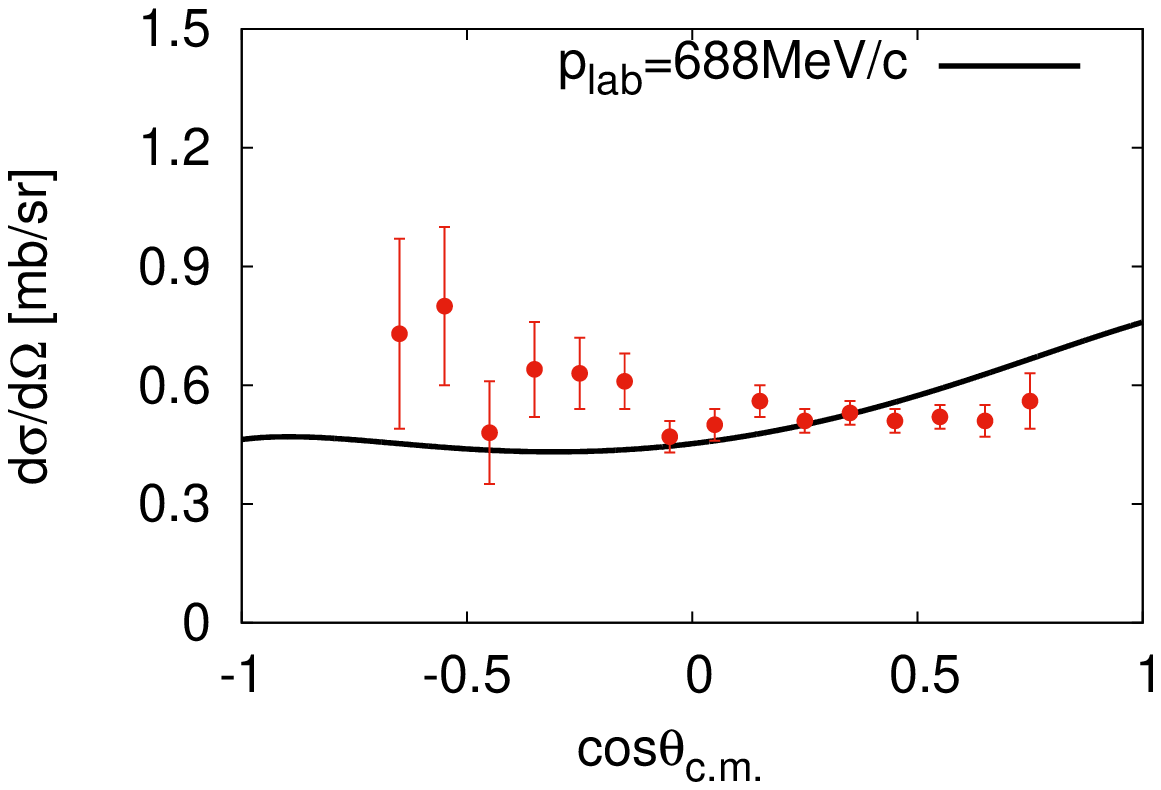}
  \end{center}
 \end{minipage}
\begin{minipage}{0.5\hsize}
  \begin{center}
   \includegraphics[width=80mm]{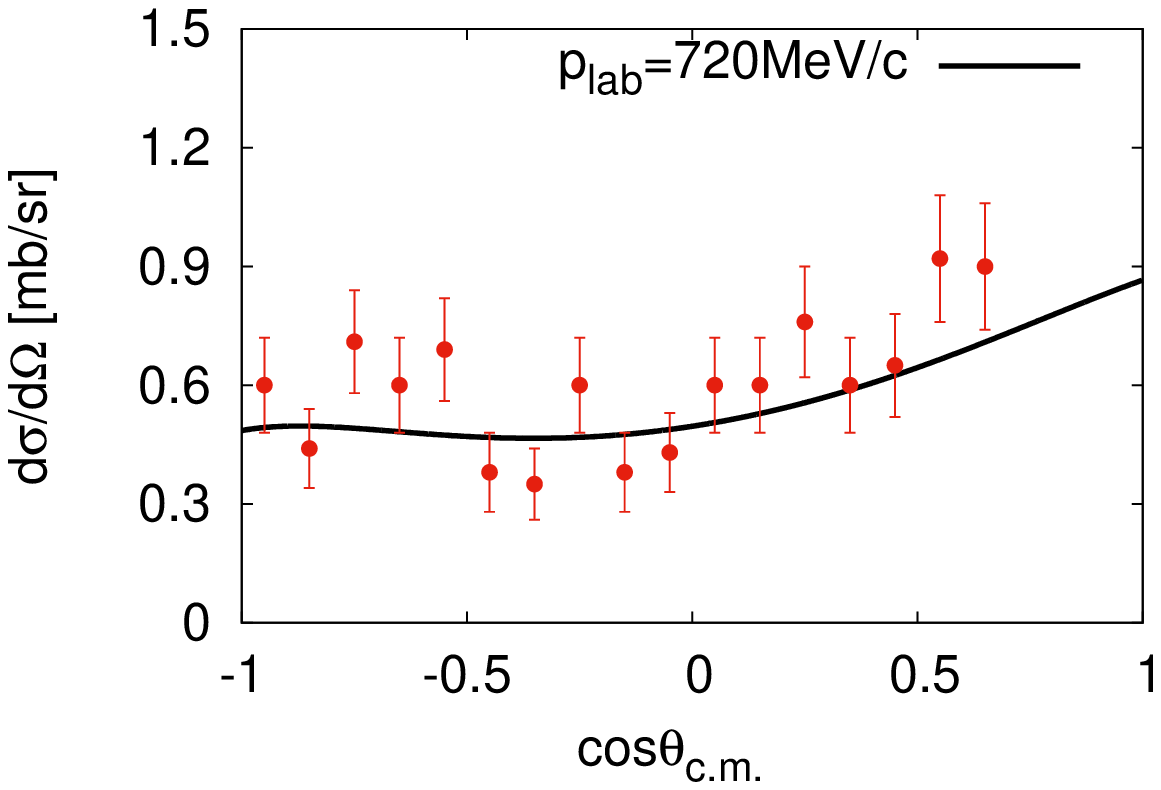}
  \end{center}
 \end{minipage}
\begin{minipage}{0.5\hsize}
  \begin{center}
   \includegraphics[width=80mm]{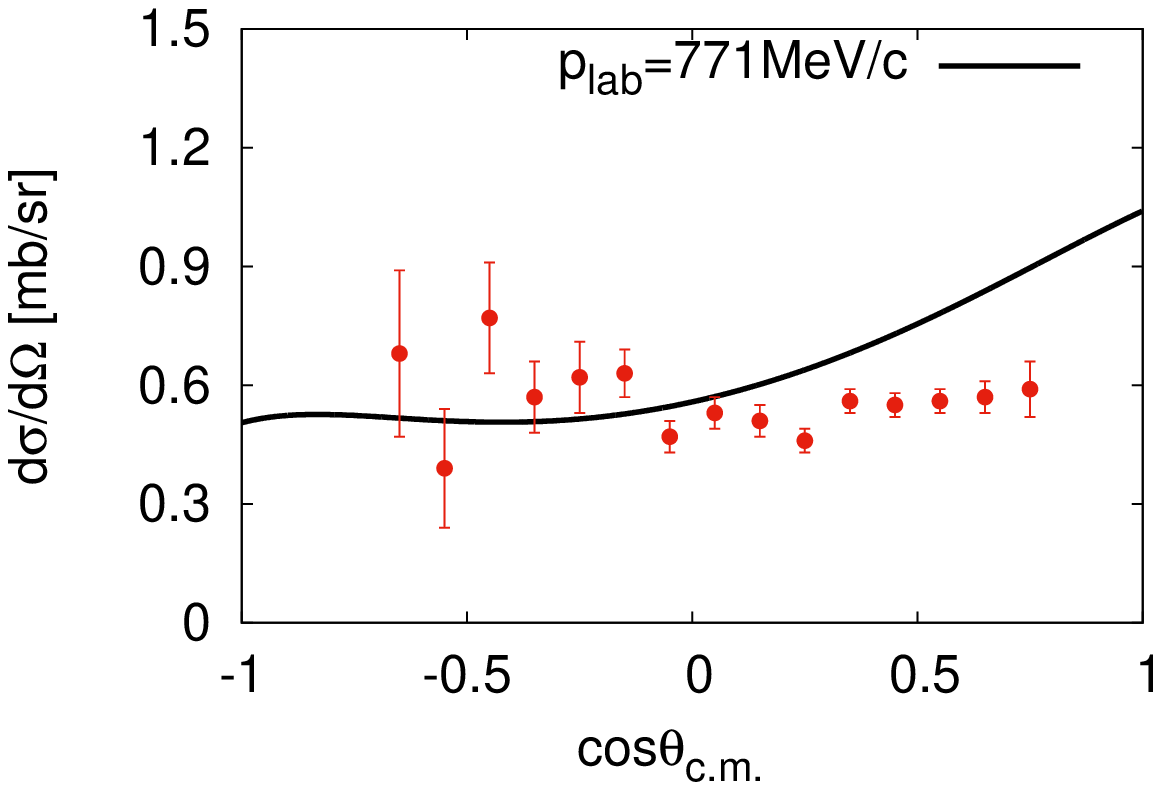}
  \end{center}
 \end{minipage}
\begin{minipage}{0.5\hsize}
  \begin{center}
   \includegraphics[width=80mm]{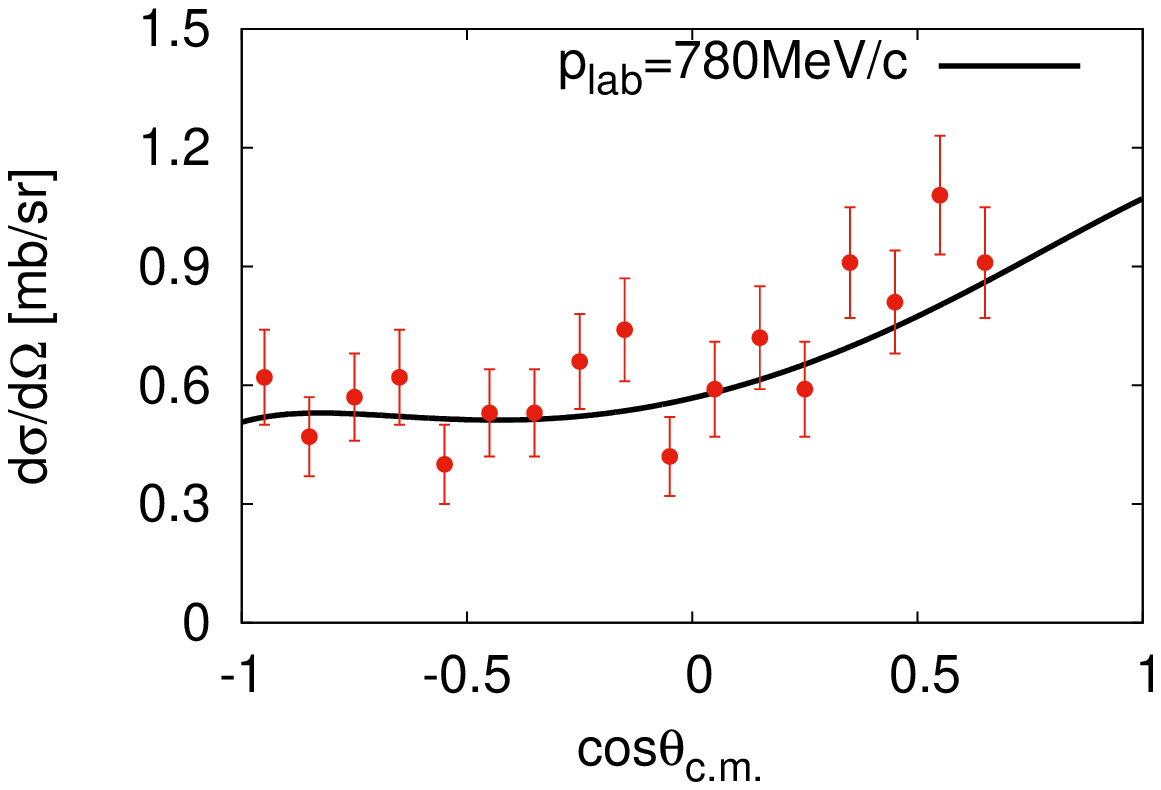}
  \end{center}
 \end{minipage}
 \vspace{-0.5cm}
 \caption{The differential cross section of $K^{+}n$ elastic scattering from unitarization
 compared with the experimental data of Ref. \cite{gia, dam1975}.
The momenta at the $p_{{\rm lab}}$=640, 720 and 780 MeV/c are the data from Ref. \cite{gia}.
The others are the data from Ref. \cite{dam1975}. }
\label{fig:kn_diff_uni}
 \end{figure}

In the following, we compare our results with the experimental values.
The results for the total cross section are shown in Figs. \ref{fig:tot1_Uni} and \ref{fig:tot0_Uni}.
As we expected, the reproduction of the data for higher momenta is better than the tree level amplitude
in Figs. \ref{fig:tot1_Tree} and \ref{fig:tot0_Tree}.
In Fig. \ref{fig:kp_diff_uni}, we show the differential cross section of the $K^{+}p$ elastic scattering together with
the experimental data \cite{cameron}. 
Although we have used only the data at $p_{\rm lab}=726$ MeV/c 
to determine the parameters, surprisingly we obtain great reproduction of the differential cross 
sections for lower energies. 
In Fig. \ref{fig:kn_diff_uni}, we show the differential cross section of the $K^{+}n$ 
elastic scattering with the experimental data taken from Refs. \cite{gia, dam1975}.
We find fairly nice reproduction, especially for $p_{\rm lab}$=640, 720 and 780 MeV/c, the agreement between 
the calculation and data is quite good both in magnitude and angular dependence. 
These data are from the same experiment \cite{gia}.  
For the other lab momenta, which are from different experiment \cite{dam1975}, 
the agreement is marginal especially for lower energies. 
We can use these data to determine 
the parameter and find a solution which provides a much better reproduction for the differential 
cross sections, but with this solution we cannot reproduce the $I=0$ total cross section.


\begin{figure}[]
 \begin{minipage}{0.5\hsize}
  \begin{center}
   \includegraphics[width=80mm]{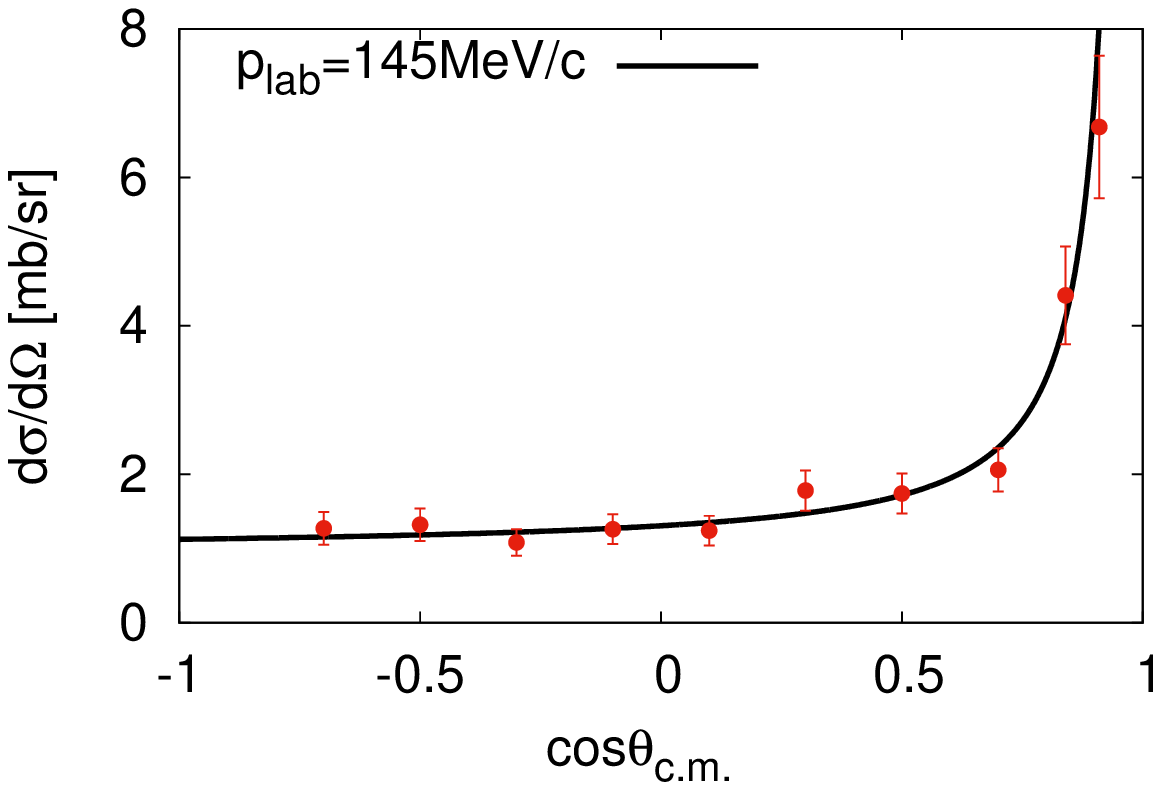}
  \end{center}
 \end{minipage}
 \begin{minipage}{0.5\hsize}
  \begin{center}
   \includegraphics[width=80mm]{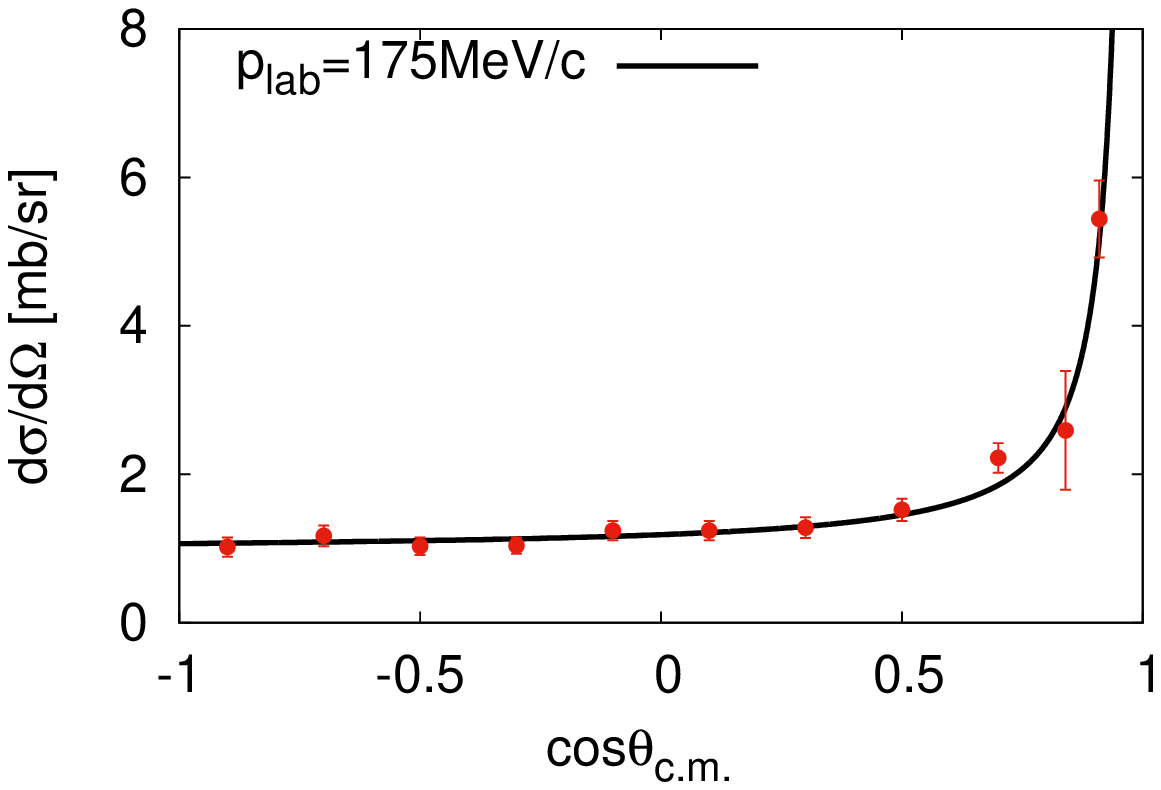}
  \end{center}
 \end{minipage}
\begin{minipage}{0.5\hsize}
  \begin{center}
   \includegraphics[width=80mm]{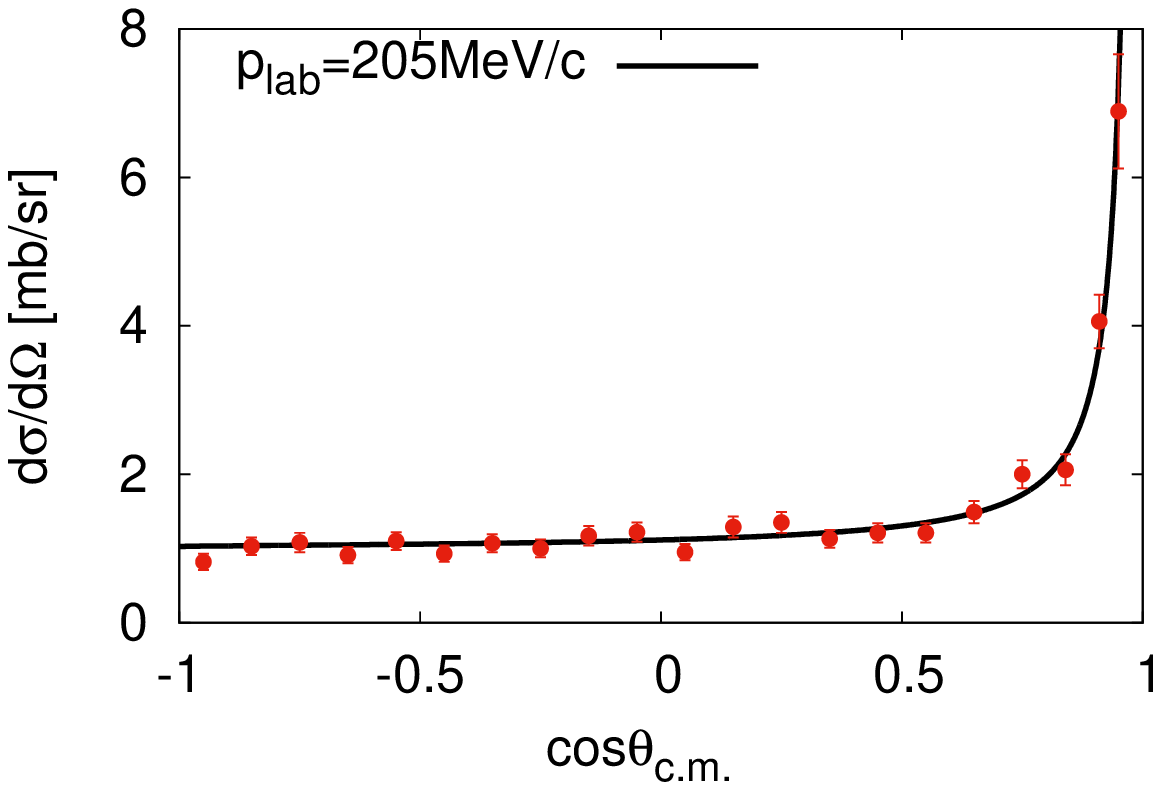}
  \end{center}
 \end{minipage}
\begin{minipage}{0.5\hsize}
  \begin{center}
   \includegraphics[width=80mm]{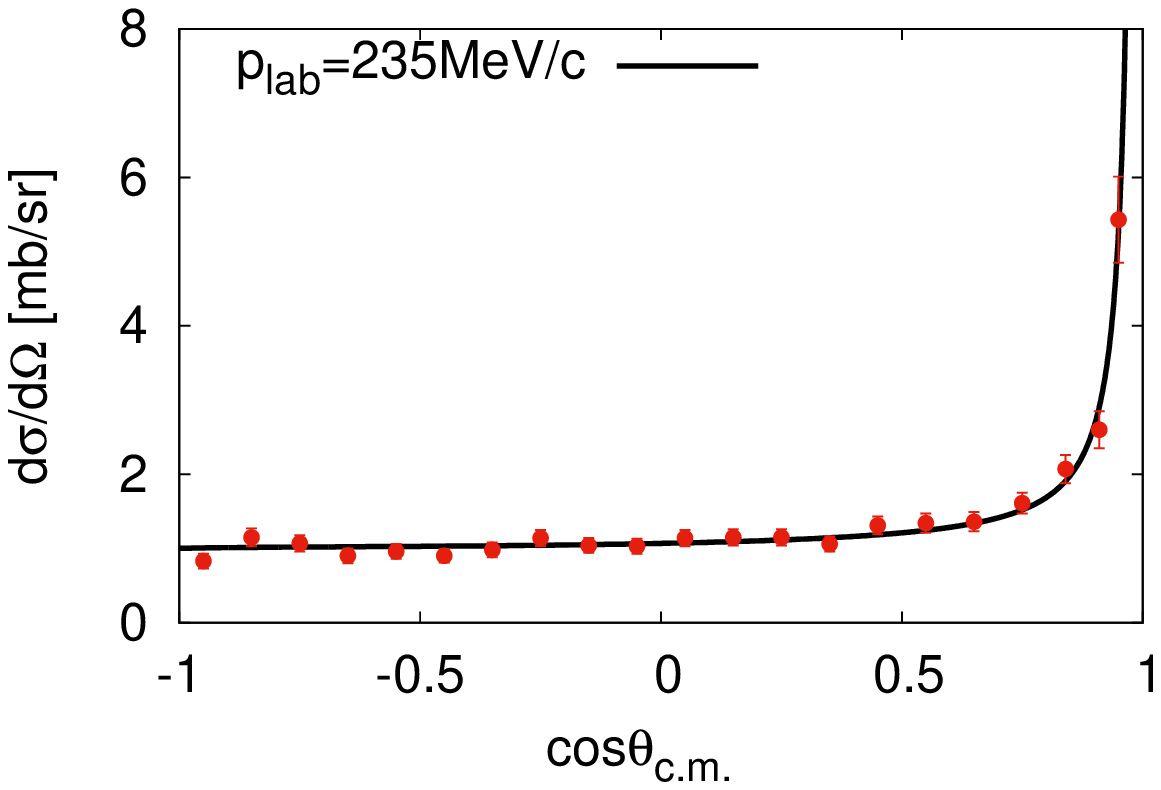}
  \end{center}
 \end{minipage}
 \begin{minipage}{0.5\hsize}
  \begin{center}
   \includegraphics[width=80mm]{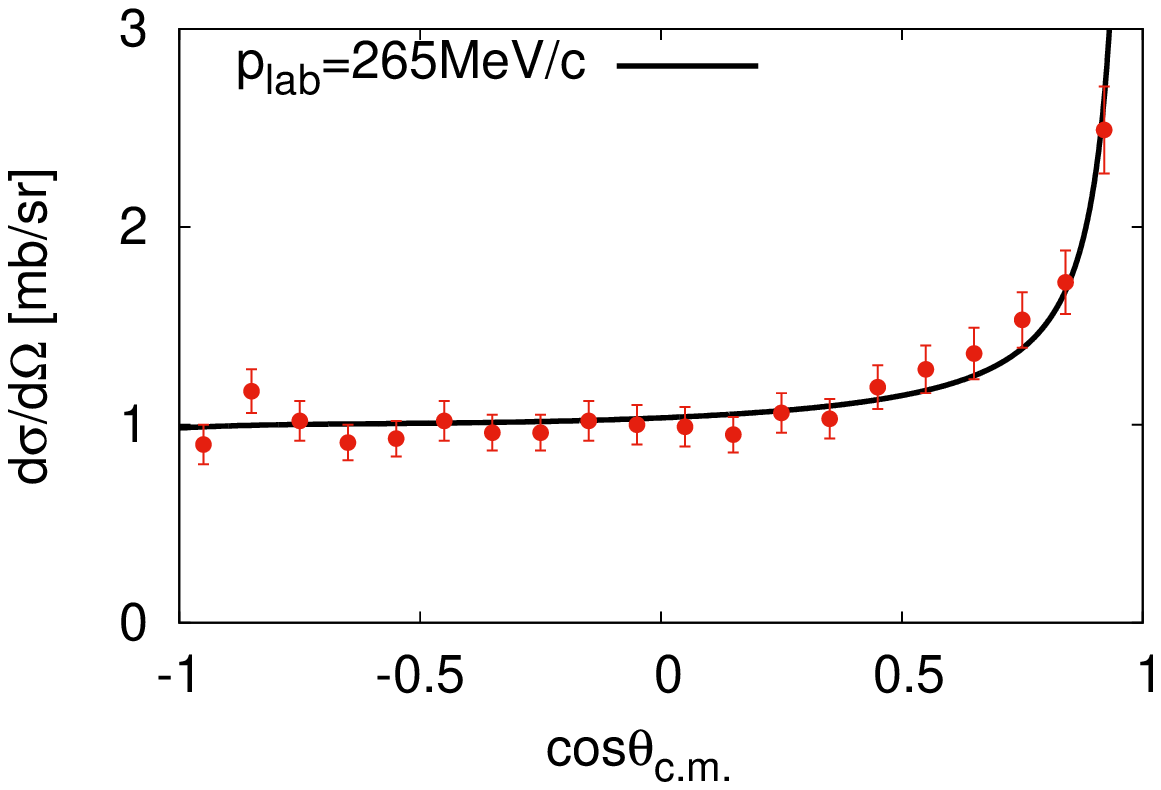}
  \end{center}
 \end{minipage}
\begin{minipage}{0.5\hsize}
  \begin{center}
   \includegraphics[width=80mm]{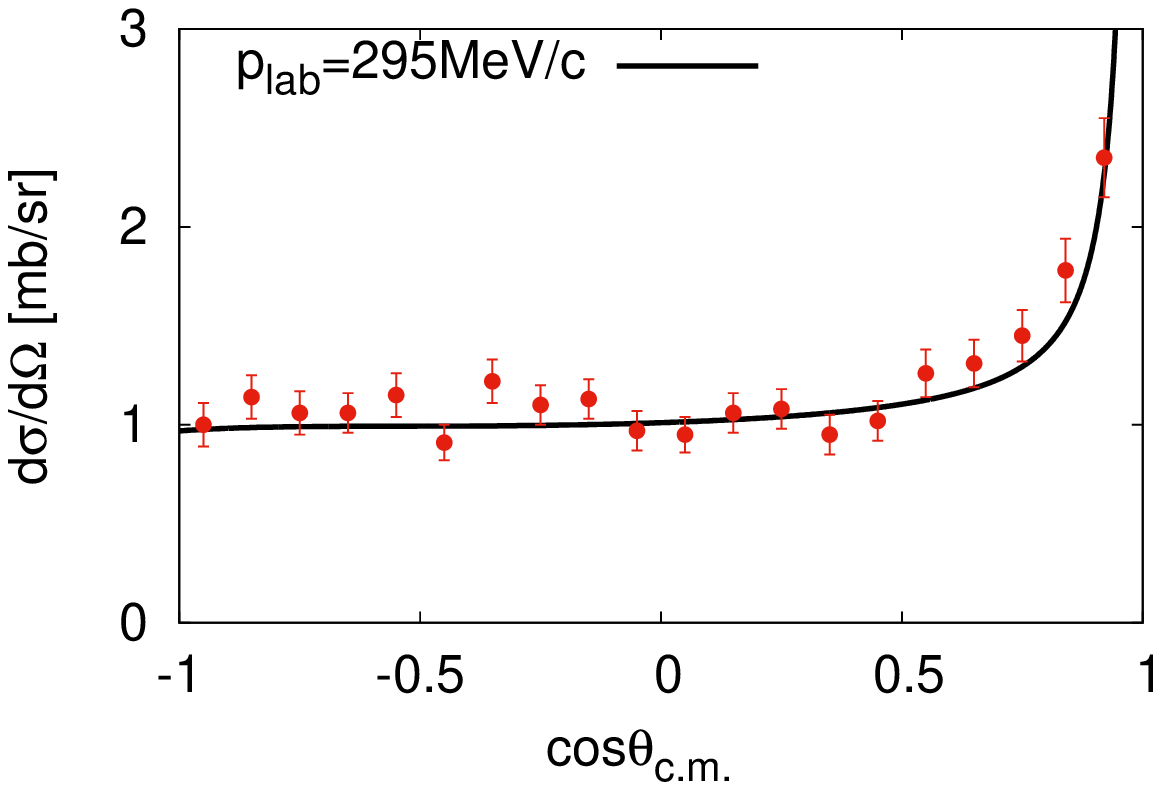}
  \end{center}
 \end{minipage}
\begin{minipage}{0.5\hsize}
  \begin{center}
   \includegraphics[width=80mm]{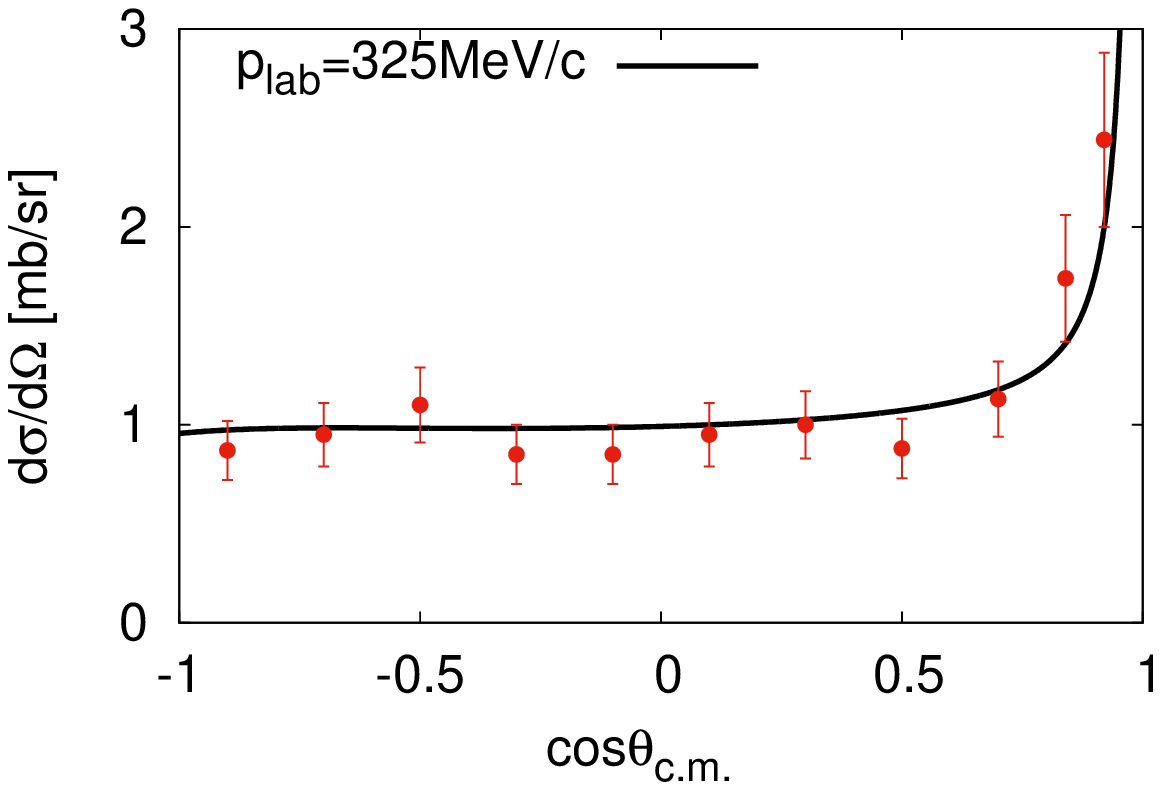}
  \end{center}
 \end{minipage}
\begin{minipage}{0.5\hsize}
  \begin{center}
   \includegraphics[width=80mm]{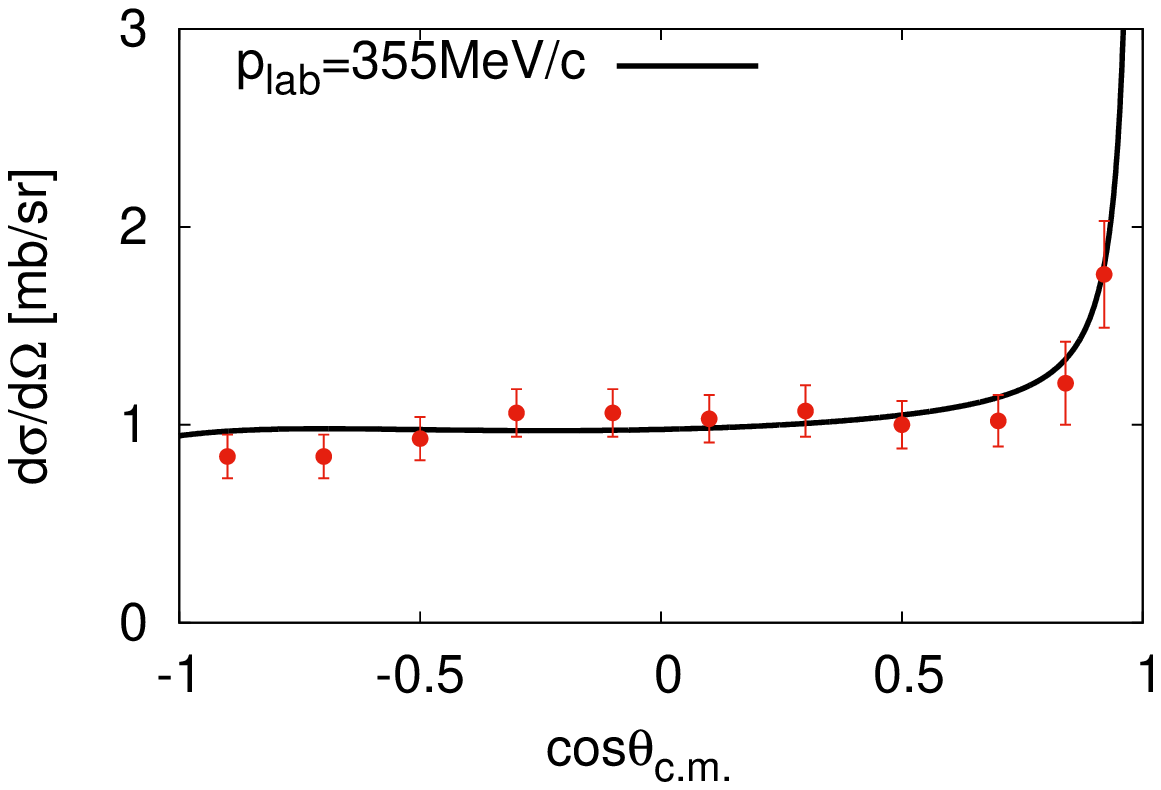}
  \end{center}
 \end{minipage}
 \end{figure}

\begin{figure}[]
 \begin{center}
  \includegraphics[width=100mm]{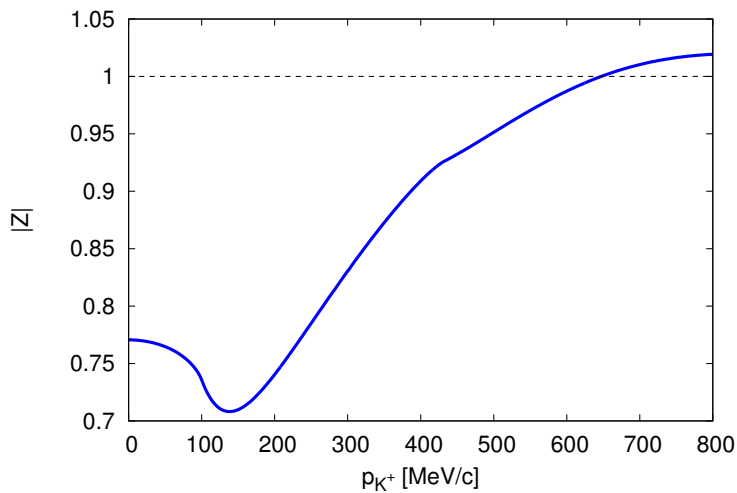}
 \vspace{-0.5cm}
 \caption{The kaon momentum dependence of the wavefunction renormalization factors $Z$
 with Fermi motion obtained by the unitarized amplitude.}
 \label{fig:comparison}
 \end{center}
\end{figure}

\begin{figure}[]
 \begin{center}
  \includegraphics[width=100mm]{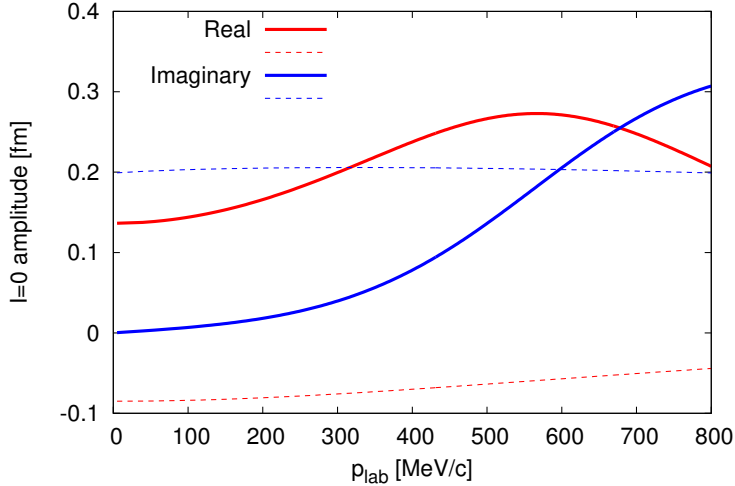}
 \vspace{-0.5cm}
 \caption{The unitarized amplitude of the $S$-wave $KN$ scattering in the $I=0$ channel
 as the function of kaon lab momentum $p_{{\rm lab}}$.
The solid lines stand for the real and imaginary parts of the original amplitude, while the dashed lines
are those of the amplitude which the pole contribution is subtracted from.}
 \label{fig:ampi0}
 \end{center}
\end{figure}

Now, we calculate the wavefunction renormalization factor $Z$ for the unitarized amplitude.
The results with Fermi motion are found in the Table. \ref{tab:z_uni} and Fig. \ref{fig:comparison}.
In the middle column of Table \ref{tab:z_uni}, $Z$ has imaginary part.
The imaginary part of the wavefunction renormalization comes from the loop contribution in the scattering amplitude. 
In our model, we consider only the $KN$ channel in the calculation of the $KN$ amplitude. 
Thus, the imaginary part comes from opening the $KN$ channel in the intermediate state. 
In the self-energy, this contribution represents the change of the nucleon state in the nucleus. 
Thus, this is a part of the breakup of the nucleus. 
Contrary to our expectation, in low momenta, our calculation shows 
that the magnitude of the wavefunction renormalization factor 
at the saturation density gets smaller than unity as seen in Table \ref{tab:z_uni}.
In Fig. \ref{fig:comparison}, we show the $K^{+}$ momentum dependence of 
the wavefunction renormalization calculated by the unitarized amplitude. 
As seen in the Fig. \ref{fig:comparison}, the wavefunction renormalization 
obtained by the unitarized amplitude drastically depends on the $K^{+}$ momentum 
and has some structure. 
In particular it crosses unity around $p_{K^{+}}=600$ MeV/c. 
Because the wavefunction renormalization factor is calculated 
by derivative of the scattering amplitude, 
the fact that $Z$ has nontrivial energy dependence implies that the scattering amplitude also 
should have nontrivial structure. 
In Fig. \ref{fig:ampi0}, we show the unitarized amplitude of
the $S$-wave $KN$ scattering in the $I=0$ channel. 
This figure shows that the imaginary 
part has rapid increase around $p_{\rm lab} = 565$ MeV/c, which correspond to 
$E_{\rm c.m.} = 1590$ MeV, and the real part has a peak at the same point. 
One naturally expects that this kind of structure stems from a pole of the amplitude in the complex energy plane. 
Actually we look for a pole of the $S$-wave amplitude with $I=0$ in the second Riemann sheet of the complex energy plane, and we find a pole at $E_{\rm c.m.}=1589 -207i$ MeV. 
In the case of the ideally narrow resonance, 
the imaginary part of the scattering amplitude should show a peak at the resonance point 
and the real part should vanish there. 
In the present case, however, because the resonance has a large width, 
the resonance strongly interfere with the scattering state. 
Consequently, due to the large interference, the roles of the imaginary and real parts are inverted. 
In Fig. \ref{fig:ampi0}, we also plot the scattering amplitude from which we have subtracted the pole contribution in dashed lines. 
One finds that the subtracted amplitude is almost flat and does not provide strong structure. 
Thus, we see that the structure in the original scattering amplitude stems from the resonance pole. 

The pole of the scattering amplitude may represent a hadronic resonance 
with spin-parity $J^{{\rm p}}={\frac{1}{2}}^{-}$, strangeness $S=+1$ and isospin $I=0$. 
This resonance is dynamically generated by the $KN$ scattering 
with the subtraction constant close to the natural value $a=-1.024$ for $\mu=1$ GeV~\cite{Hyodo:2008xr}, 
which means that the resonance is constructed almost purely by the hadronic components, 
the $K$ meson and the nucleon, without other components such as quark components. 
This resonance is certainly different from the so-called pentaquark suggested by the LEPS group~\cite{Nakano:2003qx}, 
because our resonance has a very large width. 
The narrow width pentaquark has also been discussed in the $K^{+}$-nucleus cross section  
in Refs. \cite{penta1, penta2}.
We find also that the driving force to form the $KN$ resonance is attractive contribution 
from the next-to-leading order terms, while the leading order Weinberg-Tomozawa term 
is null for the $KN$ channel with $I=0$. Hence, this resonance 
is not contradict to the argument of Refs.~\cite{Hyodo:2006yk,Hyodo:2006kg}, 
in which it is concluded that exotic channels cannot produce 
dynamical resonance of meson and baryon. 
The details of this resonance will be discussed somewhere~\cite{Aoki:2017}. 

The wavefunction renormalization obtained by the unitarized amplitude explains only a few percent of the enhancement in the optical potential around $p_{K^{+}} > 600$ MeV/c. The unitarized amplitude might not be realistic to describe the $K^{+}N$ scattering. 
Nevertheless, what we have learn in this study is that if the wavefunction renormalization factor crosses unity 
at a certain momentum, the scattering amplitude should have strong energy dependence, 
which could stem from the presence of a resonance. 
Thus, because the wavefunction renormalization factor has direct information of the scattering amplitude, 
observing the momentum dependence of the $K^{+}$ wavefunction renormalization factor 
in nuclear matter could provide a hint of the existence of an exotic hadron with $S=+1$.

 \begin{table}[]
\begin{center}
\caption{The wavefunction renormalization factor with the Fermi motion $Z$
obtained by the unitarized amplitude.}
\begin{tabular}{| c ||  c | c | c | c | } 
\hline
$p_{K^{+}}$ [MeV/c]&  $Z$&  $|Z|$   \\ \hline \hline
0.0    &                          $0.77$&    0.77                                                                  \\
130.0&                          $0.71-i0.04$&  0.71    \\
488.0&                          $0.93-i0.17$&  0.95      \\    
531.0&                          $0.95-i0.17$&  0.97         \\
656.0&                          $1.00-i0.13$&   1.01          \\
714.0&                          $1.01-i0.11$&   1.02         \\
\hline
\end{tabular} 
\label{tab:z_uni}
\end{center}
\end{table}

\section{Summarry and Conclusion}
\label{sec:5}
We have calculated the wavefunction renormalization 
to explain a part of the ``missing" repulsion of the $K^{+}N$ interaction in-medium.
This study is one of the first steps to reveal the in-medium $K^{+}N$ interaction in the context of 
the partial restoration of chiral symmetry.
The wavefunction renormalization is a next-to-leading order correction on the $K^{+}$ self-energy
in the density expansion, but the leading order correction on in-medium $K^{+}N$ scattering.
In this paper, we have done the following procedure to calculate the wavefunction renormalization;
we have calculated
the in-vacuum $K^{+}N$ interaction in chiral perturbation theory up to the
next-to-leading order.
We have determined 
the low energy constants by carrying out the $\chi^{2}$ fits 
so as to reproduce the $K^{+}N$ scattering cross section. 
Using the constructed $K^+N$ amplitude, 
we have calculated the wavefunction renormalization
by taking the energy-derivative of the $K^{+}$ self-energy 
in the nuclear medium.
The self-energy is calculated in the Thomas-Fermi approximation.
We have obtained quite good $K^{+}p$ amplitude which reproduces the differential cross sections below $p_{\rm lab}=500$ MeV/c, 
although the fitting was performed only at $p_{\rm lab} = 205$ MeV/c. 
The reproduction of the $I=0$ total cross section is also satisfying. 

We have found that the wavefunction renormalization factor at the saturation density is about 2 to 6\% depending on the kaon momentum. 
This implies that the $K^{+}N$ interaction gets 2 to 6\% enhancement in nuclear matter. 
It shows that the wavefunction renormalization explains a part of the enhancement of the $K^{+}N$ interaction in nuclear medium
and hence is one of the important medium effects for the in-medium $K^{+}$. 
We have also found that the wavefunction renormalization factor gets decreasing when the kaon momentum increases.

We also have carried out the unitarization of the amplitude obtained by chiral perturbation theory.
We have determined 
the low energy constants and subtraction constant by carrying out the $\chi^{2}$ fits 
so as to reproduce the observable. 
We have obtained the better reproduction of the $I=0$ and 1 total and differential cross section 
in the higher energies compared with the tree amplitude.
The behavior of the wavefunction renormalization obtained by the unitarized amplitude 
is quiet different to the one obtained by the tree level amplitude.
In the low energies, the wavefunction renormalization factor has the values below unity.
This behavior might be responsible for the dynamically generated resonance of the $S$-wave $KN$ scattering
in the $I=0$ channel.
Investigating the momentum dependence of the wavefunction renormalization factor could reveal the existence of an $S=+1$ resonance.

As the future prospects, we will take into accounts other in-medium effects than the wavefunction renormalization, 
such as mass modification and vertex correction.
Such effects are systematically taken into accounts by considering the in-medium chiral perturbation theory.
For this purpose, we will need to
construct a theoretical framework of the 3-flavor in-medium chiral perturbation theory.
With this approach, we will reveal the behavior of the wavefunction renormalization 
when other in-medium effects are considered.

\section*{Acknowledgment}
The work of D.J.\ was partly supported by Grants-in-Aid for Scientific Research from JSPS (25400254, 17K05449). \\

\appendix
\section{Unitarization}
\label{sec:appendix}
In this appendix, we carry out the unitarization of the scattering amplitude calculated
by chiral perturbation theory.
By carrying out the unitarization, we can obtain scattering amplitude as a complex function
consistent with elastic unitarity and we can extend energy applicable region.
For the unitarization, one performs 
non-perturbative algebraic summation of specific diagrams
using the tree level chiral perturbation theory as a potential kernel
so as to satisfy the elastic unitarity.
We work out in the isospin basis in which we decompose the $K^{+}N$ amplitude
into the isospin 0 and 1, and
carry out the unitarization in each partial waves following the method
developed in Ref. \cite{uni2}.

We perform the unitarization for each partial wave amplitude $T_{l\pm}^{I}$ by solving the Lippmann-Schwinger equation 
\begin{equation}
  T_{l\pm}^{I} = V_{l\pm}^{I} + V_{l\pm}^{I} G T_{l\pm}^{I}.
\end{equation}
This equation can be solved as an algebraic equation after the on-shell factorization with the $N/D$ method~\cite{orb2002, oller1999}, 
\begin{eqnarray}
{\cal T}_{l \pm}^{I} &=& V_{l \pm}^{I} + V_{l \pm}^{I} G V_{l \pm}^{I} + V_{l \pm}^{I} G V_{l \pm}^{I} G V_{l \pm}^{I} + \cdots \nonumber \\
&=& \frac{V_{l \pm}^{I}}{1-V_{l \pm}^{I} G}.
\end{eqnarray} 
We make good use of the tree level amplitudes obtained with chiral perturbation theory for the interaction kernel of the scattering equation after making the partial wave decomposition of
\begin{equation}
   V^{I} = T_{\rm WT}^{I} + T_{\rm Born}^{I} + T_{(2)}^{I}.
\end{equation}
The loop function $G$ is defined as
\begin{eqnarray}
G = i \int \frac{d^{4}q}{(2\pi)^{4}} \frac{1}{(P-q)^{2} - M_{N}^{2} +i\epsilon} \frac{1}{q^{2} - M_{K}^{2} + i\epsilon}.
\end{eqnarray}
The loop function can be calculated with the dimensional regularization as
\begin{eqnarray}
G&=&\frac{1}{(4 \pi )^{2}}
\biggl \{ a(\mu) 
+\ln \frac{M_{N}^{2}}{\mu^{2}} 
+\frac{M_{K}^{2}-M_{N}^{2} +s}{2 s} \ln \frac{M_{K}^{2}}{M_{N}^{2}}  \nonumber \\
&&+\frac{\bar{q}}{\sqrt{s}} \Bigl [ \ln(s - (M_{N}^{2} - M_{K}^{2}) 
+ 2\sqrt{s} \bar{q}) + \ln \left(s + \left(M_{N}^{2} - M_{K}^{2} \right) +2\sqrt{s} \bar{q} \right) \nonumber \\
&&-\ln \left(-s + \left(M_{N}^{2} - M_{K}^{2} \right) 
+2\sqrt{s} \bar{q} \right) 
-\ln\left(-s-\left(M_{N}^{2}-M_{K}^{2}\right)
+2\sqrt{s} \bar{q} \right) \Bigl]  \biggl \}
\label{eq:loop}
\end{eqnarray}
where the parameter $a(\mu)$ is the subtraction constant evaluated 
at the renormalization scale $\mu=1$~GeV and will be determined 
so as to reproduce the experimental data. 
The nucleon and kaon masses, 
$M_{N}$ and $M_{K}$, are taken at their isospin averaged physical values. 
The kinematical parameter $\bar q$ is the magnitude of the 3-momentum in the c.m.\ system. Thus,
the unitarized amplitudes are given by
\begin{eqnarray}
&& f^{I}_{{\rm uni}}(s,\theta_{\rm c.m.}) = \sum_{l=0}^{\infty} \left[ (l+1) {\cal T}_{l+}^{I} + l {\cal T}_{l-}^{I} \right] P_{l}(\cos \theta_{{\rm c.m.}}) , \\
&&g^{I}_{{\rm uni}}(s,\theta_{\rm c.m.})= \sum_{l=1}^{\infty} \left[ {\cal T}_{l+}^{I} - {\cal T}_{l-}^{I} \right]  
\sin\theta_{{\rm c.m.}}  \frac{d P_{l}(\cos\theta_{{\rm c.m.}})}{d \cos\theta_{{\rm c.m.}}}.
\end{eqnarray}


\end{document}